\documentclass[journal]{IEEEtran}
\usepackage{titlesec}
\titlespacing\section{0pt}{2pt plus 0pt minus 0pt}{0pt plus 2pt minus 2pt}
\titlespacing\subsection{0pt}{2pt plus 2pt minus 2pt}{2pt plus 2pt minus 2pt}
\titlespacing\subsubsection{0pt}{2pt plus 2pt minus 2pt}{2pt plus 2pt minus 2pt}
\usepackage{xcolor,soul,framed} 

\colorlet{shadecolor}{yellow}
\usepackage[pdftex]{graphicx}
\graphicspath{{../pdf/}{../jpeg/}}
\DeclareGraphicsExtensions{.pdf,.jpeg,.png}

\usepackage{tabularx}
\usepackage{array}
\usepackage{amsfonts}
\usepackage{amsmath}
\usepackage{cases}
\usepackage{xcolor} 
\usepackage{graphicx}
\usepackage[ruled,linesnumbered]{algorithm2e}
\usepackage{nicematrix}
\usepackage[skip=0ex]{caption}
\usepackage[noadjust]{cite}
\usepackage{bm}
\usepackage{subfig}
\usepackage{titlesec}

\begin{document}
\bstctlcite{IEEEexample:BSTcontrol}
\title{Harmonious Coexistence between Aloha and CSMA: Novel Dual-channel Modeling and Throughput Optimization}

\author{Wenhai Lin,~\IEEEmembership{Graduate Student Member,~IEEE,}
Xinghua Sun,~\IEEEmembership{Member,~IEEE,} Anshan Yuan, \\ and
Yayu Gao,~\IEEEmembership{Member,~IEEE}
\thanks{Wenhai Lin, Xinghua Sun, and Anshan Yuan are with the School of Electronics and Communication Engineering, Sun Yat-sen University, Shenzhen 518107, China (e-mail: linwh33@mail2.sysu.edu.cn; sunxinghua@mail.sysu.edu.cn; yuanansh@mail2.sysu.edu.cn;).}
\thanks{Yayu Gao is with the School of Electronic Information and
Communications, Huazhong University of Science and Technology, Wuhan
430074, China (email: yayugao@hust.edu.cn).}
\thanks{This work has been submitted to the IEEE for possible publication. Copyright may be transferred without notice, after which this version may no longer be accessible.}
}



\markboth{
}{Roberg \MakeLowercase{\textit{et al.}}: High-Efficiency Diode and Transistor Rectifiers}


\maketitle

\begin{abstract}
The scarcity of the licensed spectrum is forcing emerging Internet of Things~(IoT) networks to operate within the unlicensed spectrum. Yet there has been extensive observation indicating that performance deterioration and significant unfairness would arise, when newly deployed Aloha-based networks coexist with incumbent Carrier Sense Multiple Access (CSMA)-based WiFi networks, especially without proper adjustment of packet transmission times. 
Therefore, ensuring harmonious cohabitation between Aloha and CSMA networks is of paramount importance. 
How to properly tune system parameters to guarantee harmonious coexistence between these two networks, nevertheless, remains largely unexplored. To address the above open issue, this paper proposed a novel dual-channel analytical framework to characterize the throughput performance of the cohabitation between slotted Aloha and CSMA networks. To achieve harmonious coexistence, the total throughput of the coexisting network under a given desired throughput proportion is optimized by tuning the packet transmission time of CSMA nodes and transmission probabilities. The optimization results indicate that the packet transmission time of CSMA nodes should be set slightly less than that of Aloha nodes. The proposed framework is further applied to enhance the network throughput and fairness of the cohabitation of LTE Unlicensed and WiFi networks.


\end{abstract}

\begin{IEEEkeywords}
 Spectrum sharing, Carrier Sense Multiple Access~(CSMA), Aloha,  fairness, network throughput optimization.
\end{IEEEkeywords}

%
\IEEEpeerreviewmaketitle

\section{Introduction}
%
%
%
%
\IEEEPARstart{T}{he} limited supply of the licensed spectrum and the increasing demand for wireless data services are forcing numerous wireless communication networks to operate in the unlicensed spectrum~\cite{ramezanpour2023security,7582403}. Proprietary companies prefer to deploy emerging Internet of Things~(IoT) networks within the unlicensed spectrum because of its flexibility and the economy~\cite{8594702}. Many of these newly arrived networks adopt Aloha-based random access schemes, which do not sense the channel before transmitting. Examples include Long Term Evolution Unlicensed~(LTE-U) networks~\cite{LTEstd} and Long Range Radio Wide Area Networks~(LoRaWAN)~\cite{LoRastd}. 

The unlicensed spectrum has been widely occupied by the incumbent WiFi networks that adopt Carrier Sense Multiple
Access~(CSMA)-based random access schemes, which need to sense the channel status before transmitting. The CSMA-based WiFi networks have to share spectrum with the newly arrived Aloha-based networks. There has been extensive evidence indicating that such coexistence would result in substantial inter-network interference and significant unfairness~\cite{Google2015,cable2018,7136427,7496918,pang2017wi}. Therefore, it is a critical issue to alleviate inter-network interference and ensure harmonious cohabitation between Aloha-based and CSMA-based networks.




One salient feature of the coexistence networks in the unlicensed spectrum is that different networks have varying packet transmission
times depending on their service characteristics. For example, WiFi networks that use CSMA-based protocol are primarily designed for human-to-human~(H2H) communication, where users usually transmit large volumes of data requiring longer packet transmission times. In contrast, Internet of Things (IoT) networks~\cite{7962157}, which rely on machine-to-machine (M2M) communication and often use Aloha-based protocols, typically transmit smaller data packets, leading to shorter packet transmission times. 
It has been shown that packet transmission times play a pivotal role in determining inter-network interference~\cite{7496918,8057055}. 
Although packet transmission times designed based solely on individual network characteristics perform well in single-type networks, they struggle to ensure fairness and high throughput performance in coexistence networks. To achieve harmonious coexistence, the access parameters, the packet transmission time in particular, should be carefully tuned.

The key to obtaining optimal access parameters lies in the proper modeling of coexisting networks. Despite the simple concepts of Aloha and CSMA, the modeling of the cohabitation between Aloha and CSMA networks is exceptionally challenging. 
The challenge stems from the two heterogeneities of coexisting networks: differing node behaviors and varying packet transmission times, which will be explained in detail in Section.~\ref{sect:1-A}. The majority of analytical studies have concentrated on either Aloha or CSMA networks. How to model and optimize coexisting Aloha and CSMA networks under various packet transmission times remains largely unexplored. To address this open issue, we propose a novel dual-channel analytical framework for modeling and optimizing the coexistence performance of slotted Aloha and CSMA networks under various packet transmission times. Before diving into a detailed explanation of our results, let us first briefly review the previous studies.

\subsection{Modeling the Coexistence of Aloha and CSMA Networks: From Homogeneous to Heterogeneous Networks}
\label{sect:1-A}
There have been extensive studies on modeling and performance analysis of random access. Early research interests were centered on homogeneous networks, where the network is composed solely of nodes utilizing a single access scheme, i.e., either Aloha or CSMA. Depending on whether the modeling emphasis is on the aggregate channel or access behavior of each node, the analytical models can be generally categorized into two groups: channel-centric and node-centric. The first channel-centric model can date back to Abramson's seminal paper \cite{10.1145/1478462.1478502} on Aloha networks, which characterizes the aggregate traffic as a Poisson random variable. This channel-centric model has been broadly utilized and further developed for more complex CSMA networks \cite{1092768,5766254,5773640}. In particular, in \cite{5766254}, a semi-Markov model is proposed for characterizing the aggregate channel of a CSMA network.
For node-centric models, a well-known one was proposed by Bianchi in \cite{840210} for CSMA-based networks. Specifically, the backoff behavior of each node is characterized by a two-dimensional Markov chain. In \cite{dai2012unified}, an analytical framework was proposed to comprehensively analyze the performance of CSMA networks, where a discrete-time Markov renewal process was established to characterize the behavior of each Head-of-Line~(HOL) packet. 

With the growing diversity of networks operating in the unlicensed spectrum, research activities have increasingly intensified on coexisting networks. Extensive efforts have been made to analyze the performance of CSMA-based network coexistence, such as the coexistence of WiFi and Licensed Assisted Access~(LAA)-based LTE networks. Most of them extended the node-centric model of single access schemes~\cite{7496918,7342934,mehrnoush2018analytical,bitar2018coexistence,sutton2017delay,sun2020towards}. In particular, the authors in~\cite{7342934,mehrnoush2018analytical,bitar2018coexistence} modified the Bianchi model to evaluate the performance of CSMA-based coexisting networks. The throughput of the coexisting networks can be numerically obtained by solving a set of non-linear equations. The authors in \cite{sun2020towards} extended the unified framework in \cite{dai2012unified} to model the CSMA-based coexisting network, based on which explicit expressions of the maximum network throughput and corresponding system parameters were both obtained.

While substantial progress has been made in the modeling and performance analysis of CSMA-based coexistence networks, the coexistence of Aloha-based and CSMA-based networks remains largely unexplored. In such networks, differing node behaviors and varying packet transmission times introduce significant challenges for theoretical performance analysis. Specifically, CSMA nodes employ a listen-before-talk mechanism, while Aloha nodes transmit without checking the channel status. The differing node behavior results in intricate interactions between the two networks. For instance, the access of CSMA nodes may be deferred by the transmission activity of Aloha nodes due to the sensing process. In contrast, Aloha nodes, lacking sensing capabilities, can interfere with on-going CSMA transmissions. Existing channel-centric models, which primarily focus only on characterizing channel states, cannot be applied as they fail to capture the complex interactions arising among various types of nodes.

\begin{figure}[t]
  \centering
  \includegraphics[width=0.4\textwidth]{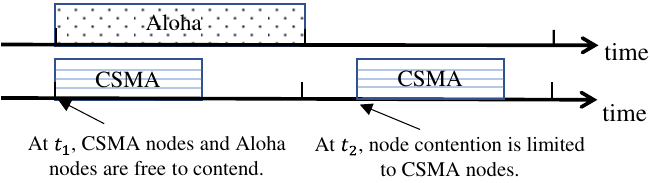}
  \caption{Illustration of node contention. A detailed explanation of this figure is provided in Section~\ref{sect:2}. }
  \label{fig:node_contention}
\end{figure}

On the other hand, in the coexistence of Aloha and CSMA networks with various packet transmission times, existing node-centric models are challenging to apply, as they typically assume a time-homogeneous probability of successful transmission. Specifically, differences in packet transmission times lead to a time-heterogeneous successful transmission probability. For example, consider that the packet transmission time of CSMA nodes is less than that of Aloha nodes. As shown in Fig.~\ref{fig:node_contention}, there exist certain periods during which the node contention is limited to CSMA nodes. The corresponding probability of successful transmission of CSMA nodes is markedly distinct from periods when both CSMA nodes and Aloha nodes are free to contend.

In one special case, i.e., where Aloha and CSMA nodes have identical packet transmission times, the probability of successful transmission becomes time-homogeneous. By assuming equal packet transmission times for both Aloha and CSMA nodes, recent work \cite{9755072} extended the node-centric approach in \cite{dai2012unified} to analyze the performance of coexisting slotted Aloha and CSMA networks. While \cite{9755072} provides valuable insights for designing coexistence networks, its assumption of identical packet transmission times limits its applicability to coexisting networks with varying transmission times. 
That calls for a clean-slate analytical framework for the coexistence of Aloha-based and CSMA-based networks with varying packet transmission times, based on which network performance, such as network throughput, can be characterized and further optimized.

\subsection{Our Contributions}
In this paper, a novel dual-channel analytical framework is proposed for modeling and optimizing the coexistence performance of slotted Aloha and CSMA networks with varying packet transmission times. In particular, we first equate the single-channel coexisting network to a dual-channel network with Aloha nodes and CSMA nodes operating on two different channels. A discrete-time Markov renewal process is established to model the channel state transition of this dual-channel network. Specifically, a three-dimensional state is introduced to describe the status of two channels and the interaction between Aloha nodes and CSMA nodes. Based on the proposed framework, the network throughputs of Aloha and CSMA can be further characterized as functions of \emph{the packet transmission time of CSMA nodes} $l_C$, network sizes, and transmission probabilities. 
The analysis indicates that the throughput of the coexisting network is closely dependent on $l_C$. With a large packet transmission time of the CSMA nodes, the throughput performance of both networks is severely degraded.

To achieve harmonious cohabitation of Aloha networks alongside CSMA networks, we further optimize the total network throughput $\hat \lambda ^{total}_{\max }$ subject to a desired throughput proportion $\gamma$ by jointly tuning system parameters. The optimization results show that for a wide range of desired throughput proportions $\gamma$ and Aloha network sizes, it is optimal to configure the packet transmission time of CSMA nodes to be slightly less than that of Aloha nodes. When the CSMA and Aloha networks aim to achieve a comparable throughput, significant throughput gains can be achieved by adopting this configuration for the packet transmission time of CSMA nodes.

A case study is provided to elaborate on how the analysis can be applied to practical coexisting LTE-U and WiFi networks and ensure their harmonious cohabitation. The simulation results demonstrate that through the joint tuning of the packet transmission time of WiFi and transmission probabilities, throughput fairness is achieved, and the total network throughput can be significantly increased compared to the sole tuning of transmission probabilities.

The remainder of this paper is organized as follows. Section \ref{sect:2} presents the analytical model. Based on which, the network throughput is characterized in Section \ref{sect:3}. Section \ref{sect:4} focuses on the throughput optimization of the coexisting network, and Section \ref{sect:5} illustrates the application of the analysis in LTE-U and WiFi coexistence. Finally, conclusions are drawn in Section \ref{sect:6}.

\begin{figure}[t]
  \centering
  \includegraphics[width=0.3\textwidth]{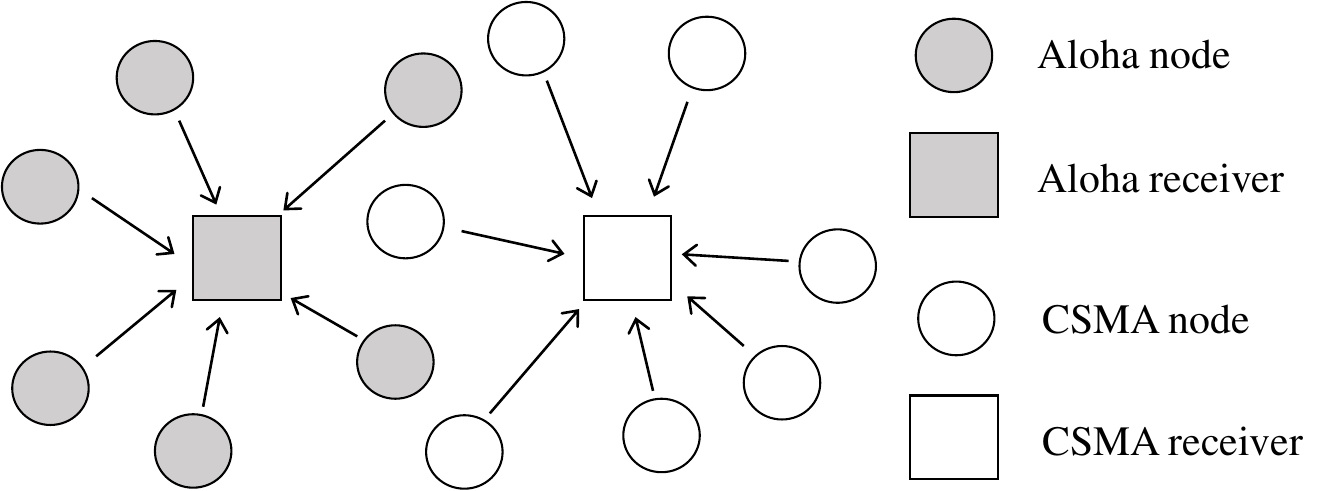}
  \caption{Graphic illustration of the Aloha networks coexisting with CSMA networks in the same spectrum.}
  \label{fig:System model}
\end{figure}

\begin{figure}[t]
  \centering
  \includegraphics[width=0.5\textwidth]{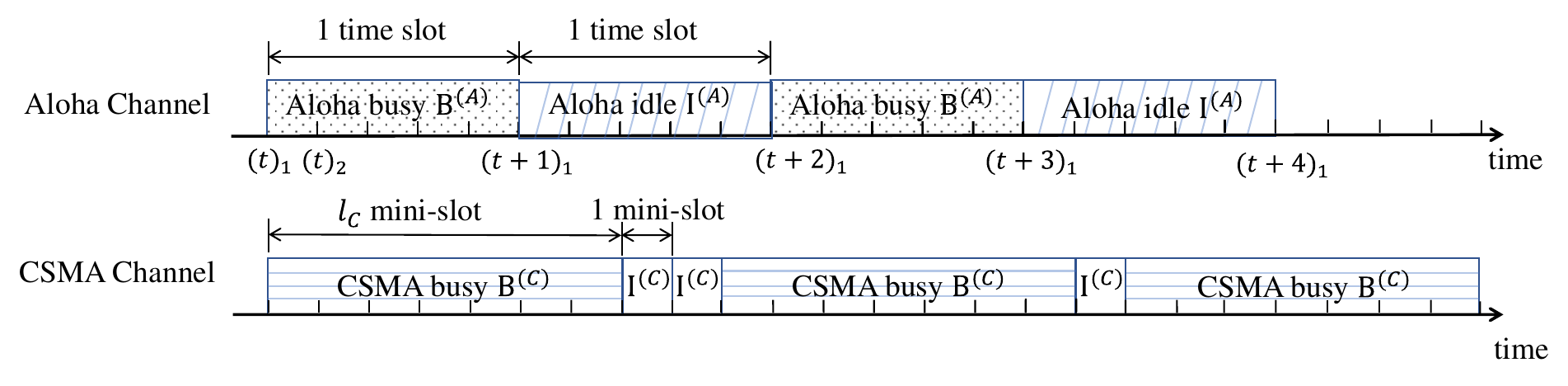}
  \caption{Illustration of the Aloha Channel and the CSMA Channel.}
  \label{fig:channel_state}
\end{figure}

\begin{figure}[t]
  \centering
  \includegraphics[width=0.35\textwidth]{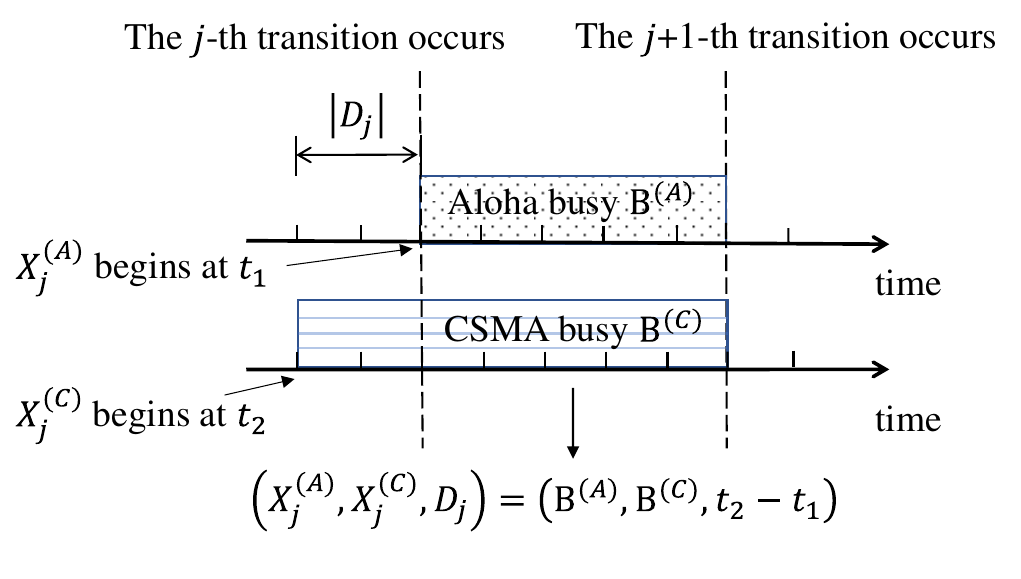}
  \caption{Illustration of definition and calculation of $D_j$.}
  \label{fig:Calculation_A}
\end{figure}

\begin{figure}[t]
  \centering
  \includegraphics[width=0.45\textwidth]{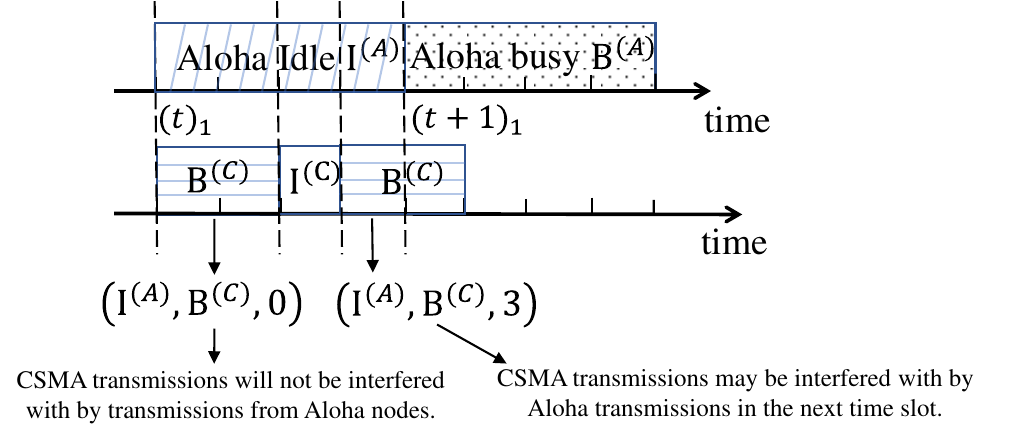}
  \caption{Illustration of function of $D_j$. $l_C=2$,~$a=0.25$.}
  \label{fig:interaction}
\end{figure}

\begin{figure}[t]
  \centering
  \includegraphics[width=0.4\textwidth]{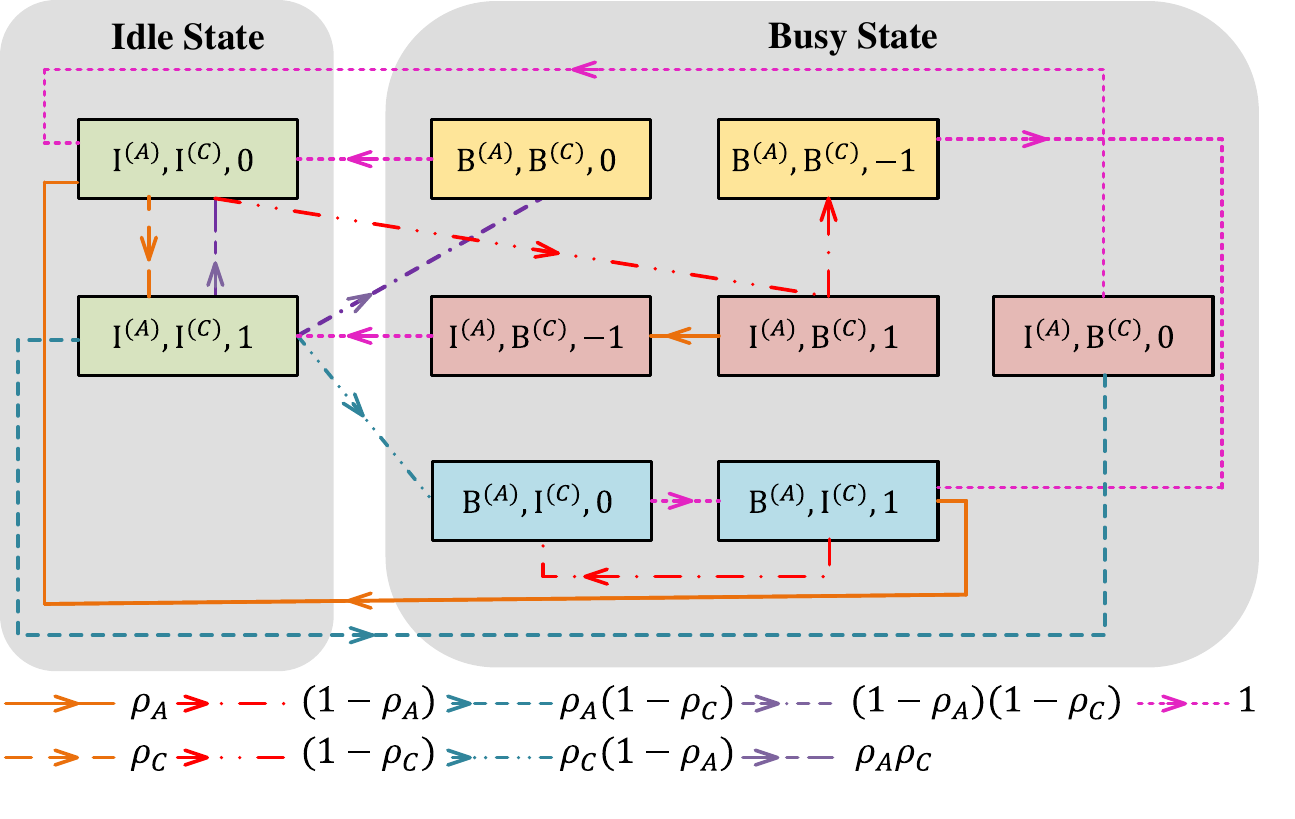}
  \caption{Embedded Markov chain $\left\{X_{j}\right\}$ given $l_C=\frac{1}{a}=2$.}
  \label{fig:Markov_chain}
\end{figure}

\section{System Model}
\label{sect:2}
Consider an $n_A$-node slotted Aloha network coexisting with an $n_C$-node slotted CSMA network. As Fig.~\ref{fig:System model} illustrates, for each network, several nodes communicate with a shared receiver. The fundamental divergence between the two networks is whether nodes perform sensing before transmitting. For each Aloha node, it would initiate transmission with a specific probability~$q_A$ when its buffer is nonempty. Conversely, each CSMA node should sense the channel. It would initiate transmission with a specific probability $q_C$ when its buffer is nonempty if it senses the channel idle, and otherwise hold its transmission. A saturated network is considered in this paper\footnote{Note that the focus of this paper is on throughput optimization, thereby placing particular emphasis on the saturated condition, where network throughput is pushed to its limit. For other performance metrics, such as delay, it is crucial to further study the coexistence performance under unsaturated conditions.}, i.e., there are always packets present in the buffers of all nodes.

As Fig. \ref{fig:channel_state} illustrates, the time axis of the Aloha network is segmented into equal intervals of unit duration called time slots. The length of one time slot is 1, equivalent to the time required for each Aloha node to transmit a single packet. On the other hand, the time axis of the CSMA network is segmented into smaller unit durations called mini-slots. Each mini-slot lasts for an interval $a$, which is equal to the ratio of the propagation delay needed by each CSMA node for sensing the channel to the time slot length, i.e., the packet transmission time of Aloha nodes. As Fig. \ref{fig:channel_state} illustrates, a time slot of Aloha is subsequently segmented into $\frac{1}{a}$ mini-slots. The $i$-th mini-slot
within a specific time slot $t$ is denoted as $(t)_i$, with $i$ ranging from 1 to $\frac{1}{a}$. Transmission initiation for Aloha and CSMA nodes is limited to the start of a time slot and a mini-slot respectively. In contrast to the assumption in \cite{9755072}, where the packet transmission time of CSMA nodes equals that of Aloha nodes, we consider that the packet transmission time of CSMA nodes $l_C$ can be \emph{any non-zero integer multiple of the mini-slot length}. The variable packet transmission time of CSMA nodes significantly complicates the throughput analysis. One of our important contributions lies in deriving explicit expressions of network throughput under different $l_C$. 

Assume that both Aloha and CSMA networks are within each other's hearing range, resulting in significant inter-network interference. 
The classical collision model\footnote{The classical collision model is a widely-adopted PHY layer assumption in the existing research studies of random access networks~\cite{10.1145/1478462.1478502,1092768,5766254,5773640,840210,dai2012unified,7342934,mehrnoush2018analytical,mehrnoush2018fairness,bitar2018coexistence,7582447,sun2020towards}. Although the classical collision model evaluates a worst-case scenario, from the interference point of view, it captures the essence of channel contention. In practice, a packet can be successfully received as long as its received signal-to-interference-plus-noise ratio (SINR) is sufficiently high, even if it overlaps with other packets.
Therefore, it is of great importance to extend the analysis to incorporate other receiver structures, such as the capture model, where a packet can be successfully transmitted as long as its received SINR exceeds a certain capture threshold.} is assumed, where one packet can be successfully received and decoded only if a unique packet transmission is active. Due to Aloha nodes without sensing capability, the transmission from Aloha nodes may partially overlap with the transmission from CSMA nodes. As depicted in Fig.~\ref{fig:channel_state}, the Aloha transmission starting at $(t+2)_1$ partially overlaps with the CSMA transmission. In this paper, if there is any non-zero overlap in packet transmissions, then these packets all fail to be decoded.

After describing the system model, let us examine Fig.~\ref{fig:node_contention}. At time $t_1$, which is the beginning of a time slot, both Aloha and CSMA nodes can participate in the transmission contention. The probability of successful transmission for a CSMA node at $t_1$ is given by $(1-q_A)^{n_A}(1-q_C)^{n_C-1}$, which is the probability that no Aloha nodes initiate a transmission and that the other $n_C-1$ CSMA nodes also do not initiate a transmission. At time $t_2$, since it occurs in the middle of a time slot and CSMA transmissions do not extend into subsequent slots, Aloha nodes do not participate in contention, leaving only CSMA nodes to compete. Consequently, the probability of a successful transmission for a CSMA node at $t_2$ is given by $(1-q_C)^{n_C-1}$. This result demonstrates that when the packet transmission time of CSMA nodes is less than that of Aloha nodes, the probability of successful transmission for a CSMA node is time-heterogeneous.

\subsection{Network Throughput}
\label{sect:2-A}

When the collision model is adopted, a maximum of one packet can be successfully received at any given time. The network throughput can be defined as the long-term fraction of the channel time used for successful packet transmissions, which reflects the access efficiency. The analysis of the throughput performance of coexisting networks is the central focus of this paper. The throughput of the Aloha network, the CSMA network, and the coexisting network are denoted as $\hat{\lambda }_{out}^{A}$, $\hat{\lambda }_{out}^{C},$ and $\hat{\lambda }_{out}^{total}$, respectively.

The long-term fraction of channel time utilized for successful transmissions corresponds to the probability the channel is in the successful transmission state~\cite{5773640}. Therefore, the critical aspect of throughput analysis resides in the characterization of the probabilities of the channel in different states. In the following, a novel dual-channel model is established to characterize the channel state transition of the Aloha and CSMA coexisting networks.

\subsection{Dual-Channel Modeling}
\label{sect:2-1}

The coexisting network can be equivalently recognized as a dual-channel network, where Aloha nodes and CSMA nodes are situated on two distinct channels—the Aloha channel and the CSMA channel, respectively. The CSMA nodes could initiate transmission only when both the Aloha channel and the CSMA channel are sensed idle (corresponding to the coexisting network is indeed idle). 

A discrete-time Markov renewal process, i.e., $\left({\bm{\mathcal{X}},\boldsymbol{V}}\right)=\left\{{\left({\boldsymbol{X}_{j},V_{j}}\right),j=0,1,\ldots}\right\}$ is established to model the state transition of the dual-channel network. Here $V_{j}$ denotes the epoch at which the $j$-th transition occurs, and $\boldsymbol{X}_{j}$ represents the state of the dual-channel network after the $j$-th transition. The state of dual-channel network $\boldsymbol{X}$ needs to describe the states of both the Aloha channel and the CSMA channel, as well as the interaction between the two types of nodes. Therefore, in this paper, the composition of $\boldsymbol{X}$ is as follows:
\begin{equation}
	\boldsymbol{X}_j=\left( X_{j}^{\left( A \right)},X_{j}^{\left( C \right)},D_{j}^{} \right),
	\label{formula2.1}
\end{equation}
where $X_{j}^{\left( A \right)}$ and $X_{j}^{\left( C \right)}$ represent the states of the Aloha channel and the CSMA channel after the $j$-th transition, respectively. The states of the Aloha channel $X^{\left( A \right)}$ are illustrated in Fig.~\ref{fig:channel_state}, which include busy Aloha channel $\mathrm{B}^{\left( A \right)}$ and idle Aloha channel $\mathrm{I}^{\left( A \right)}$ which both last for 1 time slot. The states of the CSMA channel $X^{\left( C \right)}$ include busy CSMA channel $\mathrm{B}^{\left( C \right)}$ and idle CSMA channel $\mathrm{I}^{\left( C \right)}$, which lasts for $l_C$ mini-slots and 1 mini-slot, respectively. The third term in~(\ref{formula2.1}), $D_j$, is defined as the time difference in the initiation timing of states between the CSMA channel and the Aloha channel at the $j$-th transition (in the unit of mini-slot). Fig.~\ref{fig:Calculation_A} illustrates the definition and calculation of $D_j$. 
Note that $D_j$ can well capture the interaction between two types of nodes. As shown in Fig.~\ref{fig:interaction}, given $l_C=2$ and $a=0.25$, when the network is in State $(\mathrm{I}^{(A)},\mathrm{B}^{(C)},0)$, CSMA transmissions start at the beginning of the time slot, and Aloha nodes do not transmit at $(t)_1$. Since CSMA transmissions do not extend into the next time slot, they will not be interfered with by transmissions from Aloha nodes during the entire transmission. In contrast, State~$(\mathrm{I}^{(A)},\mathrm{B}^{(C)},3)$ indicates that CSMA transmissions start at the fourth mini-slot of the time slot $(t)_4$ and will span into the next time slot. In this case, CSMA transmissions may potentially be interfered with by the Aloha transmission in the subsequent time slot.

The dimension of the embedded Markov chain $\bm{\mathcal{X}}=\left\{ \boldsymbol{X}_j \right\} $ is closely correlated with $l_C$ and $\frac{1}{a}$. In the following, we will present an example of $l_C=\frac{1}{a}=2$ to demonstrate the state transition process of $\bm{\mathcal{X}}=\left\{ \boldsymbol{X}_j \right\}$. Fig.~\ref{fig:Markov_chain} illustrates the embedded Markov chain, where $\rho_{A}=(1-q_A)^{n_A}$ and $1-\rho_{A}$ are the probabilities of no Aloha nodes initiating transmission and at least one Aloha node initiating transmission at the start of a time slot, respectively. Similarly, $\rho_C=(1-q_C)^{n_C}$ and $1-\rho_C$ are the probabilities of no CSMA nodes initiating transmission and at least one CSMA node initiating transmission, given that both the CSMA channel and the Aloha channel are idle in the previous mini-slot, respectively.

Define $\mathbb{S}$ as the state space of $\mathcal{X}$, and use $p_{\boldsymbol{\nu},\boldsymbol{\mu}}$ to represent the transition probability of the coexisting network from the State $\boldsymbol{\nu} \in \mathbb{S}$ to the State $\boldsymbol{\mu} \in \mathbb{S}$ in one step. Note that $p_{\boldsymbol{\nu},\boldsymbol{\mu}}$ is solely determined by $\boldsymbol{\nu}$ and $\boldsymbol{\mu}$, and can be obtained using conditional probability. For example, in Fig.~\ref{fig:Markov_chain}, the probability of the transition from $(\mathrm{I}^{(A)},\mathrm{I}^{(C)},0)$ to $(\mathrm{I}^{(A)},\mathrm{I}^{(C)},1)$ is given by
\begin{equation}
\begin{aligned}
&p_{(\mathrm{I}^{(A)},\mathrm{I}^{(C)},0),(\mathrm{I}^{(A)},\mathrm{I}^{(C)},1)}=\\&\mathrm{Pr}\left\{ \boldsymbol{X}_{j+1}=(\mathrm{I}^{(A)},\mathrm{I}^{(C)},1)\left| \boldsymbol{X}_{j}=(\mathrm{I}^{(A)},\mathrm{I}^{(C)},0) \right. \right\} =\rho_C.
\end{aligned}
\end{equation}
The State $(\mathrm{I}^{(A)},\mathrm{I}^{(C)},0)$ indicates that both the CSMA channel and the Aloha channel are idle, and the coexisting network is at the first mini-slot of one time slot, where solely CSMA nodes can initiate transmissions in the next mini-slot. 
Therefore, the probability that it transits to the State $(\mathrm{I}^{(A)},\mathrm{I}^{(C)},1)$ is the probability that none of CSMA nodes initiate transmission given channel was detected as idle in the prior mini-slot, i.e., $\rho_C$.

The steady-state probability distribution for the embedded Markov chain $\bm{\mathcal{X}}=\left\{ \boldsymbol{X}_j \right\} $ can be derived by
\begin{equation} 
\begin{cases}
    \pi_{\boldsymbol{\mu}} = \sum_{\boldsymbol{\nu} \in \mathbb{S}} p_{\boldsymbol{\nu},\boldsymbol{\mu}} \pi_{\boldsymbol{\nu}}, \\
    1 = \sum_{\boldsymbol{\mu} \in \mathbb{S}} \pi_{\boldsymbol{\mu}},
\end{cases}
\label{formula2.3}
\end{equation}
where $\pi_{\boldsymbol{\mu}}$ is the steady-state probability of $\boldsymbol{\mu}$. 

The interval between successive transitions, i.e., $V_{j+1}-V_{j}$, is called the holding time in State $\boldsymbol{X}_j$~(in the unit of mini-slots). Let $\tau_{\boldsymbol{\mu} }$ denote the holding time of the State $\boldsymbol{\mu}$. For State $(\mathrm{I}^{(A)},\mathrm{I}^{(C)},D)$ and $(\mathrm{B}^{(A)},\mathrm{I}^{(C)},D)$, we have
\begin{equation}
    \tau _{(\mathrm{I}^{(A)},\mathrm{I}^{(C)},D)}=\tau _{(\mathrm{B}^{(A)},\mathrm{I}^{(C)},D)}=1.
\end{equation}
Regarding State $(\mathrm{I}^{(A)},\mathrm{B}^{(C)},D)$ and $(\mathrm{B}^{(A)},\mathrm{B}^{(C)},D)$, we have two cases: $l_C\le\frac{1}{a}$ and $l_C>\frac{1}{a}$. For $l_C\le\frac{1}{a}$, the holding time is given by
\begin{equation}
\begin{mysmalls}
\tau _{(\mathrm{I}^{(A)},\mathrm{B}^{(C)},D)}=\tau _{(\mathrm{B}^{(A)},\mathrm{B}^{(C)},D)}=\begin{cases}		\frac{1}{a}-D,        &D>\frac{1}{a}-l_C,\\	l_C,  &0<D\le \,\,\frac{1}{a}-l_C,    \\l_C+D,         &D\le 0.\\\end{cases}
\end{mysmalls}
\end{equation}
For $l_C>\frac{1}{a}$, the holding time is given by
\begin{equation}
\begin{mysmalls}
    \tau _{(\mathrm{I}^{(A)},\mathrm{B}^{(C)},D)}=\tau _{(\mathrm{B}^{(A)},\mathrm{B}^{(C)},D)}=\begin{cases}	\frac{1}{a}-D,&D\ge 0,\\	\frac{1}{a},&\frac{1}{a}-l_C\le D<0,\\	l_C+D,&D<\frac{1}{a}-l_C.\\\end{cases}\\
\end{mysmalls}
\end{equation}

Note that the limiting state probabilities of the Markov renewal process $\left({\bm{\mathcal{X}},\boldsymbol{V}}\right)$ are given by~\cite{kao2019introduction}
\begin{equation}
	{\tilde{\pi}}_{\boldsymbol{\mu}} = \frac{\pi_{\boldsymbol{\mu}}\cdot \tau_{\boldsymbol{\mu}}}{{\sum\limits_{\boldsymbol{\nu} \in \mathbb{S}}\pi_{\boldsymbol{\nu}}}\cdot\tau_{\boldsymbol{\nu}}},
	\label{formula2.4}
\end{equation}
where $\boldsymbol{\mu} \in \mathbb{S}$.

\section{Network Throughput Analysis}
\label{sect:3}
As mentioned in Section~\ref{sect:2-A}, the crux of network throughput analysis lies in the probabilities of the channel in different states, which correspond to the limiting state probabilities. In this section, we first derive the limiting state probabilities of idle states\footnote{For the sake of brevity, we refer to the states that both the Aloha channel and the CSMA channel are idle, such as $( \mathrm{I}^{\left( A \right)},\mathrm{I}^{\left( C \right)}, D)$, as \emph{idle states}, while the remaining states are referred as \emph{busy states}. } and further characterize the network throughputs of Aloha and CSMA as explicit expressions of the packet transmission
time of CSMA nodes $l_C$, the
network sizes $n_A$ and $n_C$, and the transmission probabilities $q_A$ and $q_C$.

\subsection{The Limiting State Probabilities of Idle States}
\label{sect:idle_pro}
By employing the transformation steps in Appendix~\ref{app:trans}, (\ref{formula2.3}) can be reorganized into the following matrix form, which exclusively contains the limiting state probabilities of idle states :
\begin{equation}
\label{eq:idle_eq_set}
\mathbf{y}=\mathbf{A}\mathbf{y}+\mathbf{b},
\end{equation}
where $\mathbf{y}$ is the vector of the limiting state probabilities of idle states $(\mathrm{I}^{(A)},\mathrm{I}^{(C)},D)$ and $\mathbf{b}$ is a constant vector. $\mathbf{y}$ and $\mathbf{b}$ are given by
\begin{equation}
\mathbf{y}=\left[ \tilde{\pi}_{(\mathrm{I}^{(A)},\mathrm{I}^{(C)},0)},\tilde{\pi}_{(\mathrm{I}^{(A)},\mathrm{I}^{(C)},1)},\cdots ,\tilde{\pi}_{(\mathrm{I}^{(A)},\mathrm{I}^{(C)},\frac{1}{a}-1)} \right] ^\mathrm{T} 
 \label{Y_matrix}
\end{equation}
and
\begin{equation}
\mathbf{b}=\left[ a\rho_A,0, \cdots ,0,0,0 \right] ^\mathrm{T} ,
\end{equation}
respectively. $\mathbf{A}$ is the coefficient matrix. When $l_C$ and $a$ exhibit different relationships, $\mathbf{A}$ takes on different forms. The following two cases will be discussed separately: when $l_C$ is greater than or equal to $\frac{1}{a}$, and when $l_C$ is less than $\frac{1}{a}$.
\subsubsection{$l_C$ is greater than or equal to $\frac{1}{a}$}
\label{sect:eq}
For this case, $\mathbf{A}$ can be obtained as
\\[0.6cm]
\begin{equation}
\begin{mysmall}
\label{eq:equal_A}
	\mathbf{A}=\begin{bNiceMatrix}
		h & h & \cdots & h & h & e & e & \cdots & e & e \\
		v & 0 & \cdots & 0 & 0 & z & 0 & \cdots & 0 & 0 \\
		0 & v & \cdots & 0 & 0 & 0 & z & \cdots & 0 & 0 \\
		  &   & \ddots &   &   &   &   & \ddots &   &   \\
		0 & 0 & \cdots & v & 0 & 0 & 0 & \cdots & z & 0 \\
		0 & 0 & \cdots & 0 & v & 0 & 0 & \cdots & 0 & z \\
		z & 0 & \cdots & 0 & 0 & v & 0 & \cdots & 0 & 0 \\
		0 & z & \cdots & 0 & 0 & 0 & v & \cdots & 0 & 0 \\
		  &   & \ddots &   &   &   &   & \ddots &   &   \\
		0 & 0 & \cdots & z & 0 & 0 & 0 & \cdots & v & 0 \\	
		\CodeAfter
		\OverBrace[shorten,yshift=3pt]{1-1}{1-5}{\frac{1}{a}-(l_{C}~\text{mod}~\frac{1}{a})}
	\end{bNiceMatrix}_{\left( \frac{1}{a}\times\frac{1}{a}\right) },
\end{mysmall}
\end{equation}
where $h=\left \lfloor al_{C} \right \rfloor\rho_A(\rho_C-1)$, $e= \lceil al_{C} \rceil\rho_A(\rho_C-1)$, $v=\rho_C$ and $z=\rho_A\left(1-\rho_C\right)$. Given $l_C$ and $a$, we can solve~(\ref{eq:idle_eq_set}) and obtain explicit expressions of $\tilde{\pi}_{\left(\mathrm{I}^{(A)}, \mathrm{I}^{(C)}, D\right)}$ with respect to $q_A$ and $q_C$. For instance, when $l_C=5$ and $a=0.25$, $\mathbf{A}$ is given by
\begin{equation}
\begin{aligned}
 \label{eq:notint}
    &\mathbf{A}=\\&\small \left[\begin{array}{cccc} 
\rho_A\left(\rho_C-1\right) & \rho_A\left(\rho_C-1\right) & \rho_A\left(\rho_C-1\right) & 2\rho_A\left(\rho_C-1\right) \\
\rho_C & 0 & 0 & \rho_A\left(1-\rho_C\right) \\
\rho_A\left(1-\rho_C\right) & \rho_C & 0 & 0 \\
0 & \rho_A\left(1-\rho_C\right) & \rho_C & 0
\end{array}\right].   
\end{aligned}
\end{equation}
By substituting~(\ref{eq:notint}) into~(\ref{eq:idle_eq_set}), $\tilde{\pi}_{\left(\mathrm{I}^{(A)}, \mathrm{I}^{(C)}, 0\right)}$ can be obtained as
\begin{equation}
\begin{mysmalls}
\begin{aligned}
    &\tilde{\pi}_{\left(\mathrm{I}^{(A)}, \mathrm{I}^{(C)}, 0\right)}=\\&\frac{\rho_A\rho_C^2}{4(\rho_C - 1)^3\rho_A^3 + 4\rho_C(\rho_C - 1)^2\rho_A^2 + (4\rho_C^3 - 4\rho_C^2)\rho_A + 8\rho_C^3}    
\end{aligned}    
\end{mysmalls}
\end{equation}

Note that when $l_C$ is a non-zero integer multiple of $\frac{1}{a}$, $\tilde{\pi}_{\left(\mathrm{I}^{(A)}, \mathrm{I}^{(C)}, D\right)}$ can be further derived as explicit functions of $l_C$, $a$, $q_A$, and $q_C$. In this case, $\mathbf{A}$ takes the following form:
\begin{equation}
\label{eq:equal_A_1}
	\mathbf{A}=\begin{bNiceMatrix}
		 g&g & g & g & g & g & g  \\
		f & 0 &0& \cdots & 0 & 0 & 0  \\
		0 & f & 0& & 0 & 0 & 0  \\
		  & \vdots & &\ddots &   & \vdots &  \\
		0 & 0 & 0& & 0 & 0 & 0  \\
		0 & 0 & 0&\cdots & f & 0 & 0  \\
		0 & 0 & 0& & 0 & f& 0  \\		
		\CodeAfter
	\end{bNiceMatrix}_{\left( \frac{1}{a}\times\frac{1}{a}\right) },
\end{equation}
where $g=al_C\rho_A(\rho_C-1)$ and $f=v+z=\rho_C+\rho_A-\rho_C\rho_A$. From~(\ref{eq:equal_A_1}), we have
\begin{equation}
 \begin{aligned}
&\tilde{\pi}_{(\mathrm{I}^{(A)},\mathrm{I}^{(C)},\frac{1}{a}-1)}\\ &=f\cdot\tilde{\pi}_{(\mathrm{I}^{(A)},\mathrm{I}^{(C)},\frac{1}{a}-2)}=\cdots=f^{\frac{1}{a}-1}\cdot\tilde{\pi}_{(\mathrm{I}^{(A)},\mathrm{I}^{(C)},0)},
\end{aligned}   
\end{equation}
indicating when $l_C$ is a non-zero integer multiple of $\frac{1}{a}$, the probability of the coexisting network being idle decays exponentially within a time slot. Thanks to the exponential decay relationship, $\tilde{\pi}_{\left( \mathrm{I}^{(A)},\mathrm{I}^{(C)},D \right)}$ can be further obtained as an explicit function of $l_C$, $a$, $q_A$, and $q_C$, which is given by
\begin{equation}
\label{eq:idle_pro_1222}
{\small
\begin{aligned}
\tilde{\pi}_{\left( \mathrm{I}^{(A)},\mathrm{I}^{(C)},D \right)} = \frac{f^Da\rho _A\left( 1-\rho _A \right) \left( 1-\rho _C \right)}{l_Ca\rho _A\left( 1-\rho _C \right) (1-\Phi )+\left( 1-\rho _A \right) \left( 1-\rho _C \right)},    
\end{aligned}}
\end{equation}
where $\Phi=\left(\rho_A+\rho_C-\rho_A \rho_C\right)^{\frac{1}{a}}$.
The probability that the coexisting network being idle $\alpha_C$ is thus given by
\begin{equation}
\label{eq:channel_idle_pro}
\begin{aligned}
 \alpha_C = \sum_{k=0}^{\frac{1}{a}-1 } \tilde{\pi}_{\left(\mathrm{I}^{(A)}, \mathrm{I}^{(C)}, k\right)}.
\end{aligned}
\end{equation}
where $\tilde{\pi}_{\left(\mathrm{I}^{(A)}, \mathrm{I}^{(C)}, k\right)}$ is the probability that both the Aloha channel and the CSMA channel are idle, i.e., the coexisting network is idle, at the $(k+1)$-th mini-slot. By substituting~(\ref{eq:idle_pro_1222}) into (\ref{eq:channel_idle_pro}), we have
\begin{equation}
\label{eq:channel_idle_pro_1}
\begin{aligned}
 \alpha_C = \frac{a \rho_A(1-\Phi)}{l_C a \rho_A\left(1-\rho_C\right)(1-\Phi)+\left(1-\rho_A\right)\left(1-\rho_C\right)}.
\end{aligned}
\end{equation}

Note that when $l_C=\frac{1}{a}$, i.e., the packet transmission time of CSMA nodes is equal to that of Aloha nodes,~(\ref{eq:channel_idle_pro_1}) reduces to  
\begin{equation}
\begin{aligned}
 \alpha_C =\frac{a \rho_A(1-\Phi)}{\left(1-\rho_A\Phi\right)\left(1-\rho_C\right)},
\end{aligned}
\end{equation}
which is consistent to Eq.~(13) in \cite{9755072}.

\subsubsection{$l_C$ is less than $\frac{1}{a}$}

For the case that $l_C$ is less than $\frac{1}{a}$, $\mathbf{A}$ can be obtained as 
\newline
                
\begin{equation}
\vspace{12pt}
\begin{mysmall}
\label{eq:unequal_A}
	\mathbf{A}=\begin{bNiceMatrix}
		0 & 0 & \cdots & 0 & 0 & u & u & \cdots & u & u \\
		v & 0 & \cdots & 0 & 0 & z & 0 & \cdots & 0 & 0 \\
		0 & v & \cdots & 0 & 0 & 0 & z & \cdots & 0 & 0 \\
		  &   & \ddots &   &   &   &   & \ddots &   &   \\
		0 & 0 & \cdots & v & 0 & 0 & 0 & \cdots & z & 0 \\
		0 & 0 & \cdots & 0 & v & 0 & 0 & \cdots & 0 & z \\
		w & 0 & \cdots & 0 & 0 & v & 0 & \cdots & 0 & 0 \\
		0 & w & \cdots & 0 & 0 & 0 & v & \cdots & 0 & 0 \\
		  &   & \ddots &   &   &   &   & \ddots &   &   \\
		0 & 0 & \cdots & w & 0 & 0 & 0 & \cdots & v & 0 \\	

		\CodeAfter
		\OverBrace[shorten,yshift=3pt]{1-1}{1-5}{\frac{1}{a}-l_{C}}
		\UnderBrace[shorten,yshift=3pt]{10-4}{10-9}{l_{C}+1}
	\end{bNiceMatrix}_{\left( \frac{1}{a}\times\frac{1}{a}\right) },
\end{mysmall}
\end{equation}
where $u=\rho_A\left(\rho_C-1\right)$, $v=\rho_C$, $w=1-\rho_C$, $z=\rho_A\left(1-\rho_C\right)$.~(\ref{eq:unequal_A}) indicates that when $l_C$ is less than $\frac{1}{a}$, the probability of the coexisting network being idle do not decay exponentially within a time slot, as in the case that $l_C$ is a non-zero integer multiple of $\frac{1}{a}$. In this case, an explicit expression of $ \tilde{\pi}_{\left(\mathrm{I}^{(A)}, \mathrm{I}^{(C)}, D\right)}$ cannot be obtained as a function of $l_C$, $a$, $q_A$, and $q_C$. Yet for given $l_C$ and $a$, we still can solve~(\ref{eq:idle_eq_set}) and obtain an explicit expression of $ \tilde{\pi}_{\left(\mathrm{I}^{(A)}, \mathrm{I}^{(C)}, D\right)}$ in terms of $q_A$ and $q_C$ . For instance, when $l_C=2$ and $a=0.25$, $\mathbf{A}$ is given by
\begin{equation}
\label{eq:A_lessthan_special}
    \mathbf{A}=\left[\begin{array}{cccc}
0 & 0 & \rho_A\left(\rho_C-1\right) & \rho_A\left(\rho_C-1\right) \\
\rho_C & 0 & \rho_A\left(1-\rho_C\right) & 0 \\
0 & \rho_C & 0 & \rho_A\left(1-\rho_C\right) \\
1-\rho_C & 0 & \rho_C & 0
\end{array}\right].
\end{equation}
By combining~(\ref{eq:idle_eq_set}),~(\ref{eq:channel_idle_pro}) and~(\ref{eq:A_lessthan_special}), the probability that the coexisting network being idle $\alpha_C$ can be obtained as
\begin{equation}
\begin{small}
\begin{aligned}
&\alpha_C =  -\Biggl( \rho_A \biggl( (\rho_C - 1)^3 \rho_A^2 + (-4\rho_C^2 + 5\rho_C -1)\rho_A - \rho_C^3 \biggr. \Biggr.  \Biggl. \biggl. -\rho_C^2 - \\ &2 \biggr) \Biggr) 
           \bigg/ \Biggl( 4 + 4(\rho_C -1)^4 \rho_A^2 
           + (-4\rho_C^4 + 16\rho_C^2 -16\rho_C +4)\rho_A \Biggr).
\end{aligned}
\end{small}
\end{equation}

\subsection{Network Throughput}
\label{sect:net_thp}
The network throughput can be obtained from the limiting state probabilities of busy states. Appendix~\ref{app:trans} indicates that the limiting state probabilities of busy states can be represented by the limiting state probabilities of idle states that are already characterized in Section~\ref{sect:idle_pro}.
In this section, the network throughputs for Aloha and CSMA networks, $\hat{\lambda }_{out}^{A}$ and $\hat{\lambda }_{out}^{C}$ are derived as explicit expressions of the packet transmission
time of CSMA nodes $l_C$, the
network sizes $n_A$ and $n_C$, and the transmission probabilities $q_A$ and $q_C$.. 

\subsubsection{Throughput of Aloha network}
The network throughput of Aloha corresponds to the probability of the coexisting network being in the following state: One Aloha node successfully requests to transmit, and there is no CSMA transmission for the duration of the packet transmission. This probability can be derived from the limiting probability of the State $\left(\mathrm{B}^{(A)}, \mathrm{I}^{(C)}, 0\right)$. The State $\left(\mathrm{B}^{(A)}, \mathrm{I}^{(C)}, 0\right)$ indicates that no CSMA nodes transmit at the first mini-slot of transmission by the Aloha nodes. CSMA nodes will not request during the whole transmission as the channel is sensed busy by CSMA nodes. Given that only Aloha nodes initiate the transmission, the probability of successful transmission is $\frac{n_Aq_A(1-q_A)^{n_A-1}}{1-(1-q_A)^{n_A}}$. The network throughput of Aloha can be thus written as
    \begin{equation}
    \label{eq:thp_Aloha}
        \hat{\lambda }_{out}^{A}=\tilde{\pi}_{\left(\mathrm{B}^{(A)}, \mathrm{I}^{(C)}, 0\right)}\cdot \frac{n_Aq_A(1-q_A)^{n_A-1}}{1-(1-q_A)^{n_A}}\cdot\frac{1}{a},
    \end{equation}
where $\tilde{\pi}_{\left(\mathrm{B}^{(A)}, \mathrm{I}^{(C)}, 0\right)}$ is the probability that the coexisting network is at the State $\left(\mathrm{B}^{(A)}, \mathrm{I}^{(C)}, 0\right)$. The third term of the right-hand side of~(\ref{eq:thp_Aloha}), $\frac{1}{a}$, is the length of one time slot. It accounts for the fact that once the coexisting network enters the State $\left(\mathrm{B}^{(A)}, \mathrm{I}^{(C)}, 0\right)$, the coexisting network will stay in Aloha channel busy state and CSMA channel idle state for the packet transmission time of Aloha nodes, i.e., $\frac{1}{a}$ mini-slots. Since the State $\left(\mathrm{B}^{(A)}, \mathrm{I}^{(C)}, 0\right)$ lasts for 1 mini-slot, the equation needs to be multiplied by the term $\frac{1}{a}$. 

For the case that $l_C$ is a non-zero integer multiple of $\frac{1}{a}$, the throughput of the Aloha network can be obtained as follows:
\begin{equation}
\label{eq:thp_Aloha_integer}
\hat{\lambda }_{out}^{A}=\frac{n_A\rho _A^{\frac{n_A-1}{n_A}}(1-\rho _A^{\frac{1}{n_A}})(1-\rho _A)}{l_Ca\rho _A(1-\Phi )+1-\rho_A}.
\end{equation}
    Note that when $l_C=\frac{1}{a}$, i.e., the packet transmission time of CSMA nodes is equal to that of Aloha nodes, the network throughout of Aloha is 
    \begin{equation}
\hat{\lambda}_{out}^{A, l_C=\frac{1}{a}}=\frac{n_A\rho _{A}^{\frac{n_A-1}{n_A}}(1-\rho _{A}^{\frac{1}{n_A}})(1-\rho _A)}{1-\rho_A\Phi},
    \end{equation}
which is consistent to Eq.~(23) in \cite{9755072}.

Regarding the case that $l_C$ is not a non-zero integer multiple of $\frac{1}{a}$, the throughput of the Aloha network cannot be expressed as a function of $l_C$ and $a$. However, for given $l_C$ and $a$, the network throughput of Aloha can be obtained by combining~(\ref{eq:idle_eq_set}) and~(\ref{eq:thp_Aloha}). For instance, given $l_C=2$ and $a=0.25$, the throughput of Aloha network is given by
\begin{equation}
\begin{mysmalls}
\begin{aligned}
\hspace{-4.5mm}
\hat{\lambda }_{out}^{A}=\frac{n_A\rho _A^{\frac{n_A-1}{n_A}}\left(1-\rho _A^{\frac{1}{n_A}}\right)\left( 2\rho_A\rho_C^2-2\rho_A\rho_C+1\right)}{1+\left( \rho_C-1\right)^4\rho_A^2+\left( -\rho_C^4+4\rho_C^2-4\rho_C+1\right)\rho_A}.
\end{aligned}
\end{mysmalls}
\end{equation}

\begin{figure*}[ht]
  \centering
  \hspace{-5pt}
  \subfloat[]{
  \includegraphics[width=0.42\textwidth]{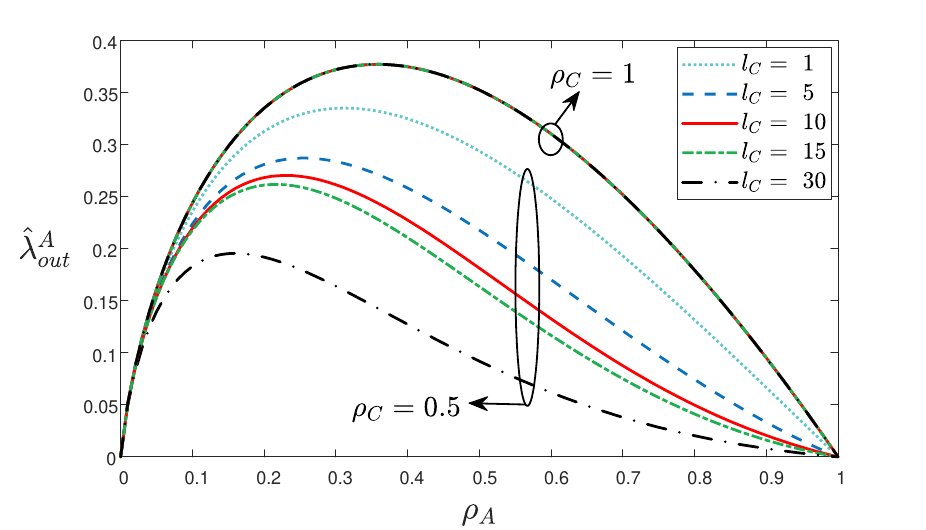}
  \label{fig:Aloha_thp_verus_lc_rouA}}
  \hspace{-25pt}
    \subfloat[]{
  \includegraphics[width=0.42\textwidth]{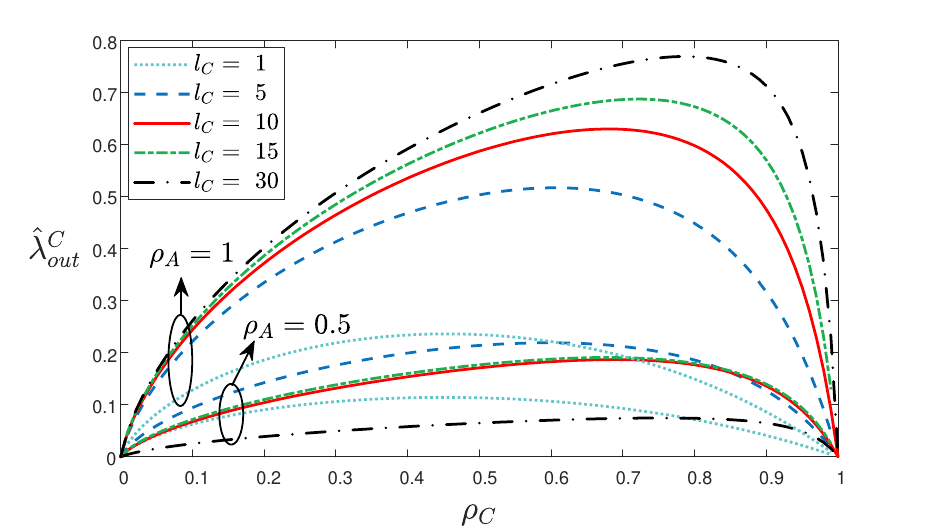}
   \label{fig:CSMA_thp_verus_lc_rouC}}
  \caption{(a) Network throughput of Aloha $\hat{\lambda }_{out}^{A}$ against $\rho_A$. $n_A=20,~l_C\in\left \{ 1,5,10,15,30 \right \},~a=0.1$. (b) Network throughput of CSMA $\hat{\lambda }_{out}^{C}$ against $\rho_C$. $ n_C=20,~\rho_A=0.5,~l_C\in\left \{ 1,5,10,15,30 \right \} ,~a=0.1$.}
\end{figure*}

Fig.~\ref{fig:Aloha_thp_verus_lc_rouA} shows how the CSMA network affects the network throughput of Aloha $\hat{\lambda }_{out}^{A}$ across various $l_C$. When $\rho_C=1$, indicating either there are no CSMA nodes~(i.e., $n_C=0$) or CSMA nodes do not attempt to transmit~(i.e., $q_C=0$), there is no interference from CSMA nodes, resulting in a high throughput for the Aloha network. Conversely, with $\rho_C=0.5$, we can see that $\hat{\lambda }_{out}^{A}$ decreases as the $l_C$ grows. This is because as the $l_C$ grows, the transmissions from CSMA nodes are more likely to extend into subsequent time slots, leading to more frequent collisions with Aloha nodes.

\subsubsection{Throughput of CSMA network}
The network throughput of CSMA can also be derived from the limiting probability of states $\left(\mathrm{I}^{(A)}, \mathrm{B}^{(C)}, D\right)$, $D\geq0$. These states indicate that Aloha nodes do not initiate transmission at the start of time slot $t$, and CSMA nodes initiate transmission at mini-slot $(t)_{D+1}$. Based on the limiting probabilities, the CSMA network throughput can then be expressed as
\begin{equation}
\begin{mysmall}
\begin{aligned}
\label{eq:thp_CSMA} 
\hat{\lambda }_{out}^{C}=&\sum_{k=0}^{\frac{1}{a}-1} \tilde{\pi}_{\left(\mathrm{I}^{(A)}, \mathrm{B}^{(C)}, k\right)}\cdot (1-q_A)^{n_A(M_k-1)} \cdot  \frac{l_C}{\tau_{\left(\mathrm{I}^{(A)}, \mathrm{B}^{(C)}, k\right)}} \cdot   \\&  \frac{n_Cq_C(1-q_C)^{n_C-1}}{1-(1-q_C)^{n_C}}.
\end{aligned}    
\end{mysmall}
\end{equation}
Note that Aloha nodes initiate transmission solely at the start of a time slot and do not perform channel sensing. Therefore, whenever CSMA transmission spans to the next time slot, the Aloha nodes may interfere with CSMA transmission midway. The second term of the right-hand side of~(\ref{eq:thp_CSMA}), $(1-q_A)^{n_A(M_k-1)}$, is the probability that Aloha nodes do not interfere with CSMA transmissions midway. $M_k$ denotes the number of time slots spanned by CSMA transmission when the CSMA transmission starts at mini-slot $(t)_{k+1}$. For $l_C$ is a non-zero integer multiple of $\frac{1}{a}$,
$M_k$ is given by
\begin{equation}
 M_k=\begin{cases}
  l_Ca,\quad& k=0, \\
  l_Ca+1,\quad& k=1,\dots,\frac{1}{a}-1.
\end{cases}   
\end{equation}
For $l_C$ is not a non-zero integer multiple of $\frac{1}{a}$, $M_k$ is given by
\begin{equation}
 M_k=\begin{cases}
  \lceil l_Ca \rceil,\quad& k=0,\dots,\frac{1}{a}-(l_C \ \text{mod}\  \frac{1}{a}), \\
  \lceil l_Ca \rceil+1,\quad& k=\frac{1}{a}+1-(l_C \ \text{mod}\  \frac{1}{a}),\dots,\frac{1}{a}-1.
\end{cases}   
\end{equation}
The third term of right-hand side of~(\ref{eq:thp_CSMA}), $\frac{l_C}{\tau_{\left(\mathrm{I}^{(A)}, \mathrm{B}^{(C)}, k\right)}}$,  accounts for the fact that once the coexisting network enters the State $\left(\mathrm{I}^{(A)}, \mathrm{B}^{(C)}, k\right)$, the transmission of CSMA nodes will last for $l_C$ mini-slots, while the State $\left(\mathrm{I}^{(A)}, \mathrm{B}^{(C)}, k\right)$ lasts for $\tau_{\left(\mathrm{I}^{(A)}, \mathrm{B}^{(C)}, k\right)}$ mini-slots. The fourth term of right-hand side $\frac{n_Cq_C(1-q_C)^{n_C-1}}{1-(1-q_C)^{n_C}}$ is the probability of only one CSMA node transmitting, given that CSMA nodes are transmitting.

Regarding the case that $l_C$ is a non-zero integer multiple of $\frac{1}{a}$, the throughput of the CSMA network can be obtained as follows:
\begin{equation}
\label{eq:thp_CSMA_integer}
\hat{\lambda }_{out}^{C}=\frac{l_Can_C\rho_A^{l_Ca+1}\rho_C^{\frac{n_C-1}{n_C}}\left(1-\rho_C^{\frac{1}{n_C}}\right)\left(1-\Phi\right)}{l_Ca\rho_A\left(1-\rho_C\right)\left(1-\Phi\right)+\left(1-\rho_A\right)\left(1-\rho_C\right)}.
\end{equation}
Note that when $l_C=\frac{1}{a}$, ~(\ref{eq:thp_CSMA_integer}) reduces to 
\begin{equation}
\hat{\lambda }_{out}^{C,l_C=\frac{1}{a}}=\frac{n_C\rho_A^2\left( 1-\rho_C^{\frac{1}{n_C}}\right)\rho_C^{\frac{n_C-1}{n_C}}\left( 1-\Phi\right)}{\left( 1-\rho_C\right)\left( 1-\rho_A\Phi\right)},
\end{equation}
which is consistent to Eq.~(26) in \cite{9755072}.

For the case that $l_C$ is not a non-zero integer multiple of $\frac{1}{a}$, the throughput of the CSMA network can be expressed as a function of $q_A$ and $q_C$ when $l_C$ and $a$ are given. For example, given $l_C=2$ and $a=0.25$, the throughput of the CSMA network can be obtained as
\begin{equation}
\small
\begin{aligned}
&\hat{\lambda}_{out}^{C} = \biggl( 
    n_{C}\rho_{A}{\rho_{C}}^{\frac{n_{C} - 1}{n_{C}}} \left( 1 - {\rho_{C}}^{\frac{1}{n_{C}}} \right) 
    \biggl( 2(\rho_{C} - 1)^{3}{\rho_{A}}^{2} 
    \Biggr. \Biggl. + \left( -2{\rho_{C}}^{3} \right. \\& 
    \left. - 2{\rho_{C}}^{2} + 3\rho_{C} - 1 \right)\rho_{A} - \rho_{C} - 1 \biggr) 
\biggr) 
\biggr/ 
\biggl( 2 + 2(\rho_{C} - 1)^{4}{\rho_{A}}^{2}+\\& \left( -2{\rho_{C}}^{4} + 8{\rho_{C}}^{2} - 8\rho_{C} + 2 \right)\rho_{A} 
\biggr).
\end{aligned}
\end{equation}

It can be observed from Fig.~\ref{fig:CSMA_thp_verus_lc_rouC} that the packet transmission time of CSMA nodes $l_C$ significantly affects the throughput of the CSMA network. When $\rho_A=1$, interference from the Aloha network is nonexistent, and the network throughput $\hat{\lambda}_{out}^{C}$ increases as the $l_C$ grows. With $\rho_A=0.5$, the network throughput $\hat{\lambda}_{out}^{C}$ reaches its high value when $l_C$ is slightly less than $\frac{1}{a}$, i.e., $l_C=5$. The network throughput $\hat{\lambda}_{out}^{C}$ is severely degraded with a large $l_C$ or a too-small $l_C$. This degradation occurs because a large $l_C$ results in the transmission of CSMA spanning multiple time slots, making them more susceptible to interference from Aloha nodes. Conversely, a too-small $l_C$ causes CSMA nodes to waste excessive time on channel sensing, leading to poor throughput performance.

From the above results, it is clear that in coexisting Aloha and CSMA networks, the inter-network interference significantly hinges on the packet transmission time of the CSMA nodes $l_C$. A large $l_C$ leads to more frequent collisions between Aloha and CSMA nodes, resulting in the degraded performance of both networks. On the contrary, a small $l_C$ minimizes the interference from CSMA transmissions to Aloha transmissions. Yet it significantly compromises the throughput performance for CSMA due to the sensing overhead. Therefore, the packet transmission time of CSMA nodes $l_C$ is a crucial parameter for mitigating inter-network interference and enhancing access efficiency, which
should be properly designed to optimize the performance of the coexisting network. In the next section, the throughput performance of the coexisting network is further optimized by properly adjusting $l_C$, $\rho_A$, and $\rho_C$.

\section{Throughput Optimization}
\label{sect:4}
\begin{figure*}[ht]
  \centering
  \hspace{-5pt}
  \subfloat[]{
  \includegraphics[width=0.42\textwidth]{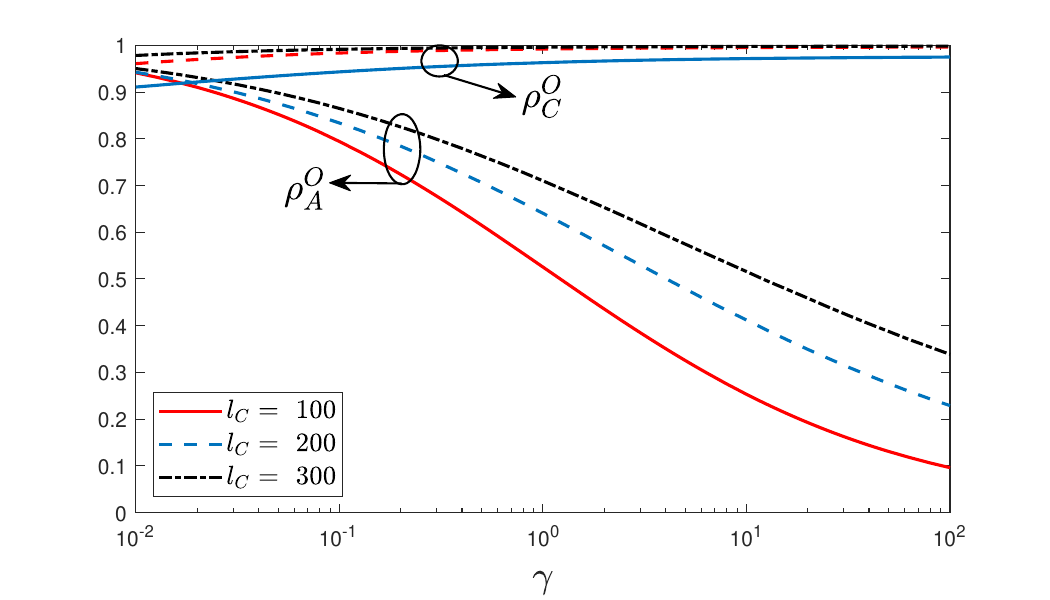}
  \label{fig:total_thp_verus_lc}}
  \hspace{-25pt}
    \subfloat[]{
  \includegraphics[width=0.42\textwidth]{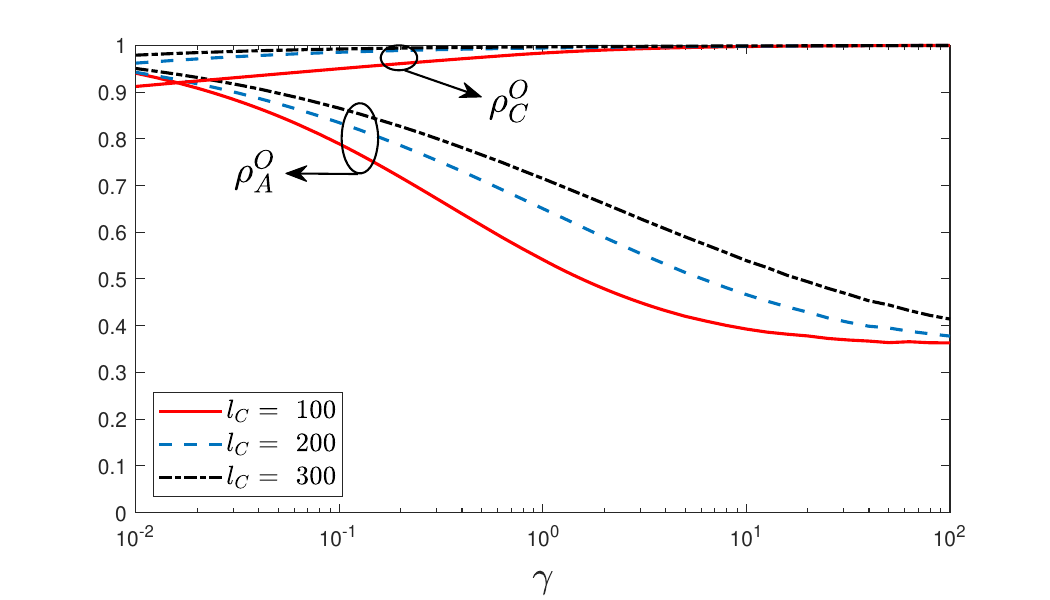}
   \label{fig:thp_optimal_lc_default_lc}}
  \caption{The optimal solutions of~(\ref{eq:opt-main}), $\rho_A^O$ and $\rho_C^O$ against desired throughput proportion $\gamma$. $l_C\in\left \{ 100,200,300 \right \} ,~a = 0.01$. (a) $n_A = 1$. (b) $n_A = 20$. }
   \label{fig:thp_optimal_solution}
\end{figure*}
The analysis in~Section \ref{sect:3} reveals that the system parameters settings, especially the packet transmission time of CSMA nodes $l_C$, are critical determinants of coexisting networks' throughput performance.
In this section, we focus on finding the optimal parameter settings to achieve harmonious cohabitation of Aloha networks alongside CSMA networks.

To optimize the coexisting network throughput performance and ensure fairness between the Aloha network and the CSMA network, we consider the throughput optimization of the coexisting network under a desired throughput proportion. In particular, the optimization problem can be written as
\begin{equation}
    \hat \lambda ^{total}_{\max }=\max _{\rho _{A},\;\rho _{C},\;l_C} \;\;\; \hat \lambda ^{A}_{out}+\hat \lambda ^{C}_{out}, \ 
    ~s.t.~\frac {\hat \lambda ^{A}_{out}}{\hat \lambda ^{C}_{out}}=\gamma,  
    \label{eq:opt-main}
\end{equation}
where $\gamma$ is the desired throughput proportion between network throughputs of Aloha and CSMA, and $\hat \lambda ^{total}_{\max}$ is the maximum total network throughput under the constraint of desired throughput proportion $\gamma$, by optimally adjusting $\rho_A$, $\rho_C$, and $l_C$.

We discuss this optimization problem in the following two cases: when $l_C$ is a non-zero integer multiple of $\frac{1}{a}$ and when $l_C$ is not a non-zero integer multiple of $\frac{1}{a}$.

\begin{figure*}[ht!]
  \centering
  \hspace{-5pt}
  \subfloat[]{
  \includegraphics[width=0.42\textwidth]{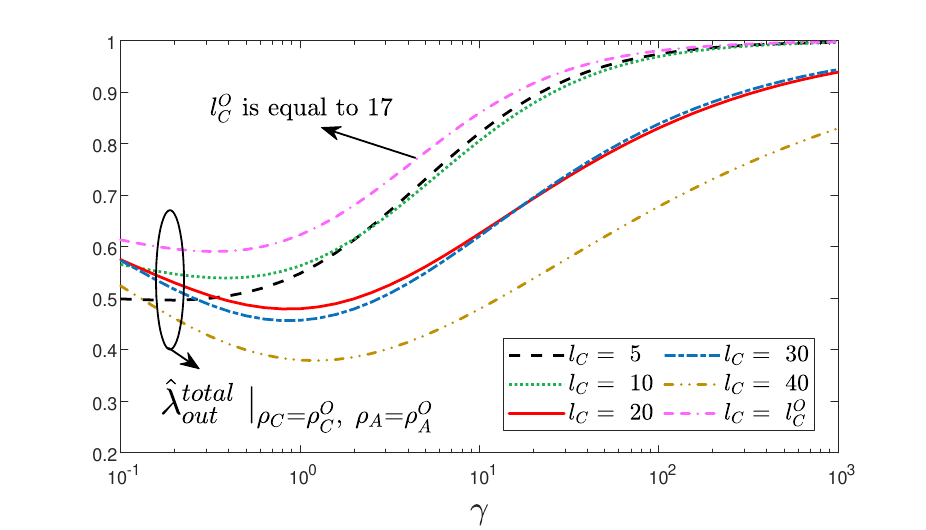}
  \label{fig:total_thp_verus_gamma_na=1}}
  \hspace{-25pt}
    \subfloat[]{
  \includegraphics[width=0.42\textwidth]{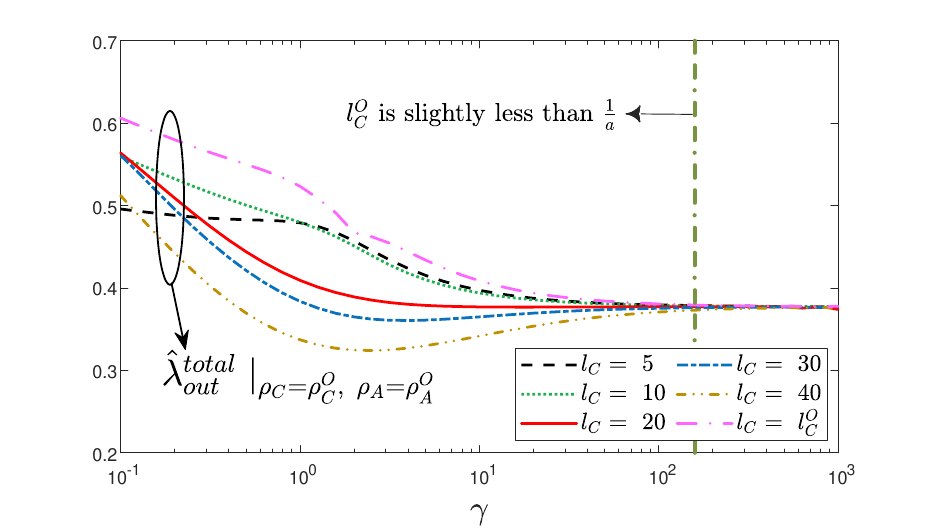}
   \label{fig:total_thp_verus_gamma_na=20}}
  \caption{Total network throughput $\hat \lambda ^{total}_{out}$ with optimal $\rho_C^O$ and $\rho_A^O$ against desired throughput proportion $\gamma$. $l_C\in\left \{ 5,10,20,30,40,l_C^O \right \},~n_C=20,~a=0.05$. (a) $n_A=1$. (b) $n_A=20$.}
    \label{fig:total_thp_verus_gamma_1}
\vspace{-20pt}
\end{figure*}

\subsection{$l_C$ is a non-zero integer multiple of $\frac{1}{a}$}
In this case, $\hat \lambda ^{A}_{out}$ and $\hat \lambda ^{C}_{out}$ can be expressed as explicit functions of $l_C$, $a$, $\rho_C$, and $\rho_A$, that is, (\ref{eq:thp_Aloha_integer}) and (\ref{eq:thp_CSMA_integer}). 
By combining~(\ref{eq:thp_Aloha}) and (\ref{eq:thp_CSMA_integer}), the optimal solutions of (\ref{eq:opt-main}) can be found through numerical methods.
Fig.~\ref{fig:thp_optimal_solution} demonstrates the optimal $\rho_A^O$ and $\rho_C^O$, as the packet transmission time of CSMA nodes $l_C$ varies from 100 to 300, and $\gamma$ shifts between 0.01 and 100. Fig.~\ref{fig:thp_optimal_solution} reveals that $\rho_C^O$ approaches 1 across a broad range of $l_C$ and $\gamma$. Based on $\rho_{C}^O\approx 1$, $\hat{\lambda}_{out}^{C}$ is approximated as
\begin{equation}
    \label{eq:temp02}
    \hat{\lambda}_{out}^{C}=\frac{l_Ca\rho_A^{l_Ca+1}\left(1-\Phi\right)}{l_Ca\rho_A\left(1-\Phi\right)+\left(1-\rho_A\right)},
\end{equation}
according to~(\ref{eq:thp_CSMA_integer}), where $\Phi=\left(\rho_A+\rho_C-\rho_A \rho_C\right)^{\frac{1}{a}}$.  
With~(\ref{eq:thp_Aloha})~and (\ref{eq:temp02}),~(\ref{eq:opt-main}) can then be simplified as
\begin{equation}
\label{eq:temp03}
\begin{aligned}
    \hat \lambda ^{total}_{\max }|_{n_{A}=1} \approx&\max _{\rho _{A},\;\rho _{C},\;l_C} \;\;\; \frac {(1+\gamma)(1-\rho _{A})\rho _{A}^{l_Ca}}{\left(1-\rho_A\right)+\rho_{A}^{l_Ca}\gamma}, \\
    &\text{s.t.}~\gamma=\frac {\left(1-\rho_A\right)^2}{l_Ca\rho_{A}^{l_Ca+1}\left(1-\Phi\right)\rho_C},\\
    &\quad \ \quad l_Ca\in \mathbb{N}^+,
\end{aligned}
\end{equation}
and
\begin{equation} 
\label{eq:temp04}
\begin{aligned}
    \hat \lambda ^{total}_{\max }|_{n_{A}\gg 1} \approx&\max _{\rho _{A},\;\rho _{C},\;l_C} \;\;\; \frac {-(1+\gamma)\rho _{A}^{l_Ca}\ln {\rho _{A}}}{\rho_{A}^{l_Ca}\gamma-\rho_A\ln{\rho_A}}, \\
    &\text{s.t.}~\gamma=\frac {-(1-\rho _{A})\ln {\rho _{A}}}{l_Ca\rho_{A}^{l_Ca}\left(1-\Phi\right)},\\
    &\quad \ \quad l_Ca\in \mathbb{N}^+. 
\end{aligned}
\end{equation}
The partial derivatives of the objective functions in (\ref{eq:temp03}) and~(\ref{eq:temp04}) with respect to $l_C$ show that the optimal packet transmission time of CSMA nodes $l_C^O$ is given by
\begin{equation}
    \label{optimal_lc}
    l_C^{O}|_{n_A=1}=l_C^{O}|_{n_A\gg1}=\frac{1}{a}.
\end{equation} For $n_A=1$, the subsequent equation's unique non-zero solution is the optimal $\rho _{A}^{O}|_{n_A=1}$
\begin{equation}
\begin{aligned}
&a\gamma \rho_{A}^{2}\mathbb{W}_{0}\left(\exp\left\{\frac{1}{a}(1-\rho_{A})\right\}\right)
\left(\!1 - (1 - \rho_{A}) \vphantom{\frac{1}{a}} \right. \\[-0.5ex]
&\quad \cdot \left(\!1 - \mathbb{W}_{0}\left(\!\exp\left\{\frac{1}{a}(1 - \rho_{A})\right\}\!\right)\!\right)\!^{\frac{1}{a}} 
= (1 - \rho_{A})^{3},
\end{aligned}   
\end{equation}
then $\rho _{C}^{O}| _{n_A=1}$ is determined as
\begin{equation} \rho ^{O}_{C}|_{n_{A}=1}=\frac {a\mathbb {W}_{0}\left ({\exp \left \{{\frac {1}{a}\left ({1-\rho ^{O}_{A}|_{n_{A}=1}}\right)}\right \}}\right)}{1-\rho ^{O}_{A}|_{n_{A}=1}}.\end{equation}
Based on this,~$\hat \lambda ^{total}_{\max }|_{n_{A}=1}$ could be approximately written as
\begin{equation} \hat \lambda ^{total}_{\max }|_{n_{A}=1}\approx \frac {(1+\gamma)(1-\rho ^{O}_{A}|_{n_{A}=1})\rho ^{O}_{A}|_{n_{A}=1}}{\rho ^{O}_{A}|_{n_{A}=1}(\gamma -1)+1}.\end{equation}
When $n_{A}\gg 1$,~$\rho ^{O}_{C}|_{n_{A}\gg 1}$ and $\rho ^{O}_{A}|_{n_{A}\gg 1}$ are
\begin{equation}
\begin{small}
 \begin{aligned}
    &\rho ^{O}_{C}|_{n_{A}\gg 1} =\\&1+a\ln \hspace {-1mm}\left ({\!\!1+\frac {1-\sqrt {1+\frac {4}{\gamma }}}{2}\hspace {-0.5mm}\left ({\!\exp \left \{{\!\frac {\sqrt {\gamma ^{2}+4\gamma }-\gamma }{2}}\right \}-1\!\!}\right)\!\!}\right) \\&/\,\left ({1-\exp \left \{{\frac {1}{2}\left ({\gamma -\sqrt {\gamma ^{2}+4\gamma }}\right)}\right \}\hspace {-1mm}}\right)
\end{aligned}   
\end{small}
\end{equation}
and
\begin{equation} \rho ^{O}_{A}|_{n_{A}\gg 1}=\exp \left \{{\frac {1}{2}\left ({\gamma -\sqrt {\gamma ^{2}+4\gamma }}\right)}\right \},\end{equation}
respectively, based on which $\hat \lambda ^{total}_{\max }|_{n_{A}\gg 1}$ could be approximately written as
\begin{equation}
\begin{small}
 \begin{aligned}
    &\hat \lambda ^{total}_{\max }|_{n_{A}\gg 1} \\&\approx\frac {(1+\gamma)\left ({\sqrt {\gamma ^{2}+4\gamma }-\gamma }\right)\exp \left \{{\frac {1}{2}\left ({\gamma -\sqrt {\gamma ^{2}+4\gamma }}\right)}\right \}}{\sqrt {\gamma ^{2}+4\gamma }+\gamma }.
    \end{aligned}   
\end{small}
\end{equation}

\subsection{ $l_C$ is not a non-zero integer multiple of $\frac{1}{a}$}In this case, $\hat \lambda ^{A}_{out}$ and $\hat \lambda ^{C}_{out}$ cannot be expressed as explicit functions of $l_C$ and $a$. Therefore, we will demonstrate the optimal packet transmission time of CSMA nodes $l_C^O$ found by the numerical method.

Fig.~\ref{fig:total_thp_verus_gamma_1} illustrates the total network throughput $\hat \lambda ^{total}_{out}$ with optimal $\rho_C^O$ and $\rho_A^O$, under three different $l_C$ settings, including $l_C$ greater than or equal to $\frac{1}{a}$~($l_C=20,30,40$), $l_C$ less than $\frac{1}{a}$~($l_C=5,10$), and $l_C$ being the optimal setting~($l_C=l_C^O$). Fig.~\ref{fig:total_thp_verus_gamma_na=1} shows the scenario with just a single Aloha node, that is, $n_A=1$. As Fig.~\ref{fig:total_thp_verus_gamma_na=1} illustrates, when $0.1\leq\gamma\leq1000$, the optimal packet transmission time of CSMA nodes $l_C^O$ equals $17$, which is slightly less than $\frac{1}{a}$. The rationale behind this is that a large $l_C$ can enhance the access efficiency of CSMA nodes~\cite{dai2013toward}, but also leads to severe inter-network interference. Specifically, the setting where $l_C$ is slightly less than $\frac{1}{a}$ can enhance access efficiency compared to the setting where $l_C$ is much less than $\frac{1}{a}$, and mitigate the interference of CSMA nodes on Aloha nodes compared to the setting where $l_C \ge\frac{1}{a}$. Note that for
the setting where $l_C\ge\frac{1}{a}$, CSMA transmissions are much more likely to extend to the next time slot compared to the setting where $l_C<\frac{1}{a}$, and potentially collide with Aloha nodes’ transmissions in the next time slot.

In Fig.~\ref{fig:total_thp_verus_gamma_na=1}, we can observe that when $l_C>=\frac{1}{a}$, the total network reaches the lowest point when $\gamma \approx 1$. This occurs because, when either CSMA or Aloha nodes have limited transmission opportunities (i.e., when $\gamma \gg 1$ or $\gamma \ll 1$), inter-network interference remains minimal, resulting in high total network throughput. However, when CSMA and Aloha nodes are required to achieve identical throughput ($\gamma = 1$), strong inter-network interference arises, reducing total network throughput. This highlights the trade-off between maximizing throughput and maintaining fairness. However, \textit{by optimally adjusting the packet transmission time, this trade-off can be significantly improved}. As shown in Fig.~\ref{fig:total_thp_verus_gamma_na=1}, when $l_C=l_C^O$, the total network throughput at $\gamma=1$ does not experience significant loss. This occurs because, with the optimal packet length configuration, inter-network interference can be significantly reduced.

Fig.~\ref{fig:total_thp_verus_gamma_na=20} illustrates the scenario where there are 20 Aloha nodes, that is, $n_A\gg 1$. As depicted in Fig.~\ref{fig:total_thp_verus_gamma_na=20}, when $0.1\leq\gamma <100$, the optimal packet transmission time of CSMA nodes $l_C^O$ is slightly less than $\frac{1}{a}$. 
When $\gamma \geq100$, the throughput gaps among different $l_C$ settings are minimal, considerably smaller than the throughput gaps when $n_A=1$. This is because when $n_A\gg 1$, the Aloha network faces not only inter-network interference but also intra-network interference. The latter cannot be alleviated by tuning $l_C$. 

Combining Fig.~\ref{fig:total_thp_verus_gamma_na=1} and Fig.~\ref{fig:total_thp_verus_gamma_na=20}, we can conclude that the setting where $l_C$ is slightly less than $\frac{1}{a}$ is the optimal $l_C$ setting for throughput performance across a broad range of $\gamma$ and $n_A$. In particular, when aiming for the CSMA network and the Aloha network to reach a comparable throughput, i.e., $\gamma=1$, the coexisting network can obtain significant throughput gain from tuning $l_C$ to be slightly less than $\frac{1}{a}$ compared to default setting where $l_C$ equals $\frac{1}{a}$. Since the inter-network interference is strong at this point, tuning $l_C$ to mitigate inter-network interference has a noticeable effect on improving the total network throughput.

\section{Case Study: Harmonious Cohabitation of LTE-U and WiFi networks}
\label{sect:5}
In the previous sections, we analyze the cohabitation performance of general Aloha and CSMA networks. The Aloha and CSMA access schemes are both widely adopted by practical wireless communication networks, such as LTE-U and WiFi networks. In this section, we will demonstrate how the preceding analysis can be applied in the cohabitation of LTE-U and WiFi networks.

One primary method for LTE networks to operate in the 5GHz unlicensed bands is duty-cycle-based LTE, also known as LTE-U~\cite{wang2017survey}.
The duty-cycle-based LTE schedules transmissions according to a specified duty cycle and is oblivious to the channel status when a transmission is scheduled to start. In contrast, the incumbent WiFi in unlicensed bands adopts CSMA-based Distributed Coordination Function~(DCF), which needs to sense channel status before transmission. When these two networks with distinctly different access schemes coexist in the same bands without properly tuning the system parameters, significant performance degradation and unfairness in the coexisting network would occur. A recent work~\cite{9755072} studied how to jointly tune system parameters to optimize the total throughput of the coexisting LTE-U and WiFi networks under a fairness constraint. However, due to the limitations of the analytical model adopted in~\cite{9755072}, the packet transmission time of WiFi was assumed to be equal to a subframe of LTE-U in~\cite{9755072}. In practice, the packet transmission time depends on the packet length and significantly impacts the coexistence network's performance. Yet how to tune the packet transmission time of WiFi nodes to optimize the coexisting network throughput under a fairness constraint on a desired throughput proportion remains largely elusive, which is the primary subject of this section.

\begin{figure*}[ht]
  \centering
  \includegraphics[height=0.15\textwidth]{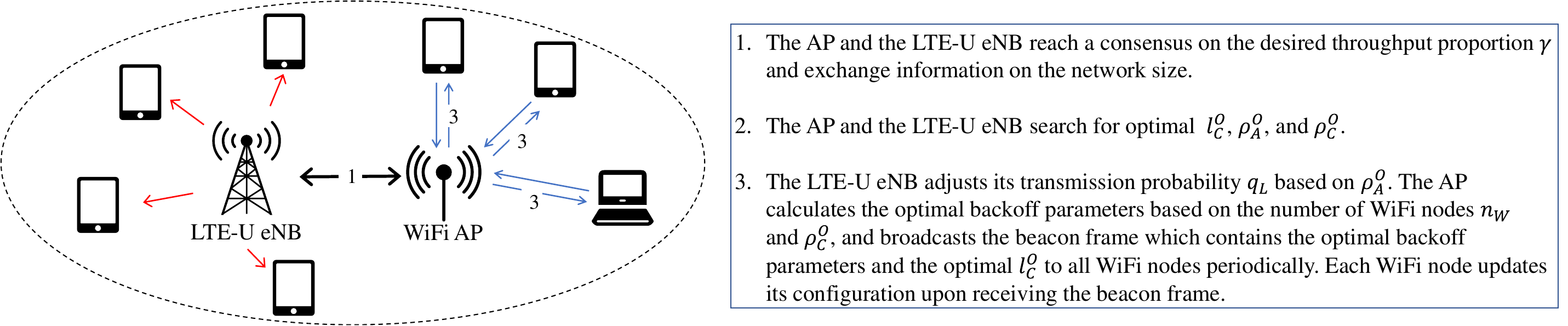}
  \caption{Illustration of the coexistence scenario~(left) and the optimal configuration implementation~(right).}
  \label{fig:Aloha_CSMA_imple}
  \vspace{-20pt}
\end{figure*}
\begin{figure}[t]
  \hspace{5pt}
\includegraphics[width=0.4\textwidth]{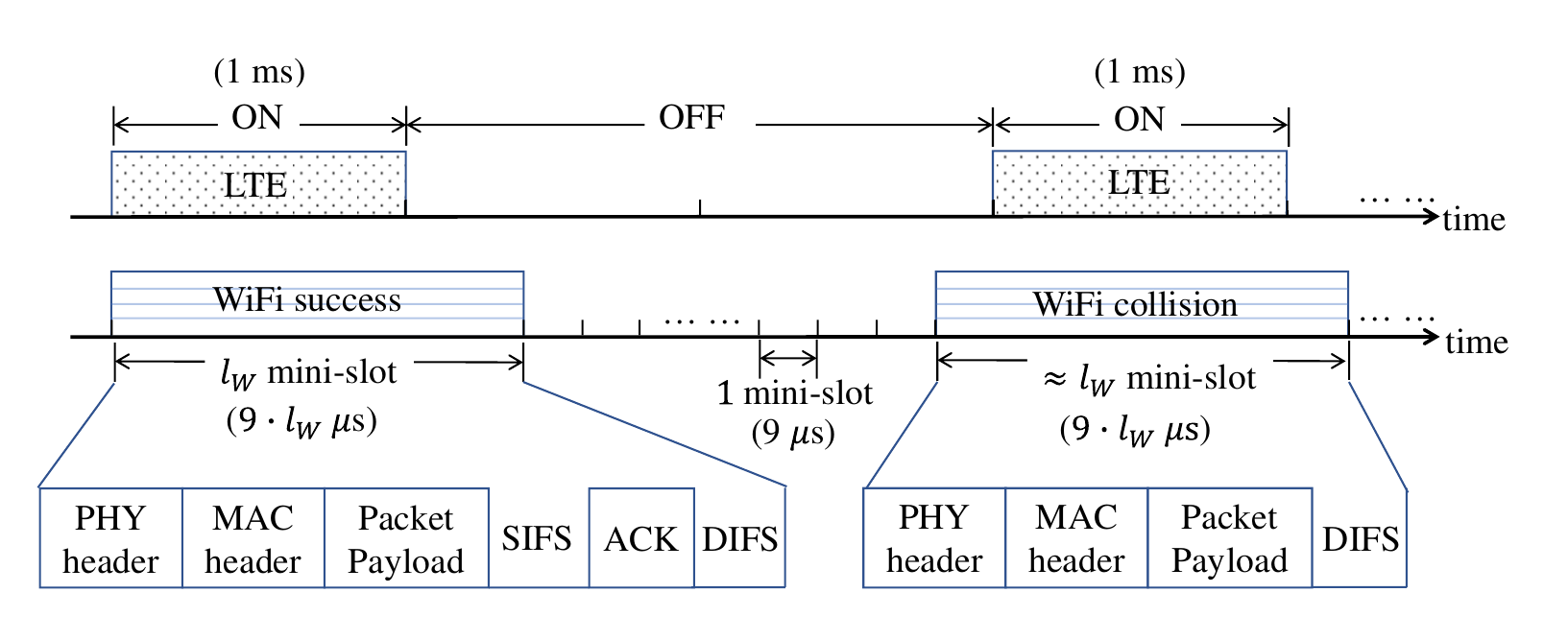}
  \caption{Illustrating transmissions from LTE-U and WiFi networks.}
  \label{fig:frame_structure}
\end{figure}
\begin{figure}[t]
\vspace{-10pt}
 \includegraphics[width=0.45\textwidth]{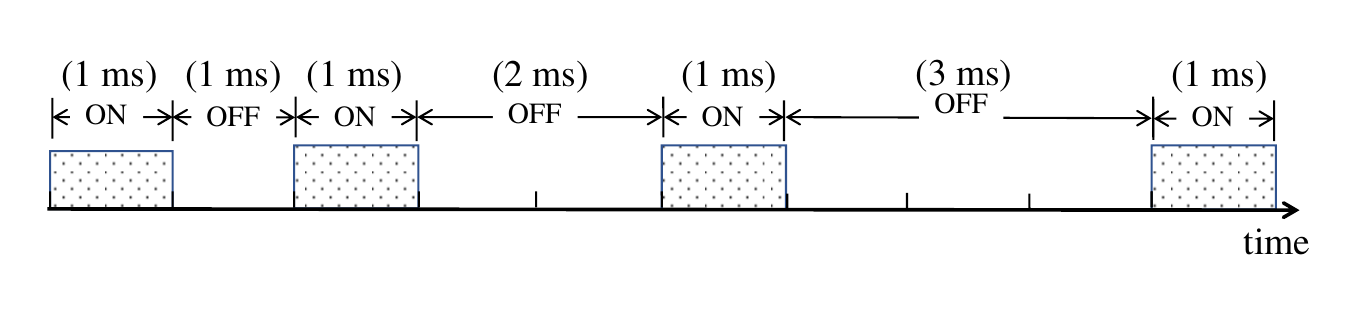}
 \vspace{-10pt}
  \caption{Illustrating transmission scheduling for the LTE-U eNB following a randomized OFF period duration.}
  \label{fig:random_OFF}
\end{figure}


\begin{figure*}[ht!]
  \centering
  \hspace{-5pt}
  \subfloat[]{
  \includegraphics[width=0.42\textwidth]{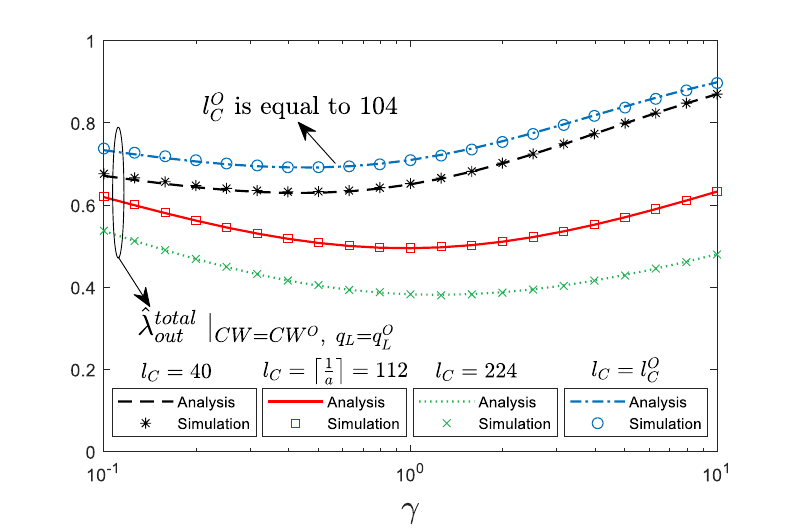}
  \label{fig:maxthp_vs_gamma_lc}}
  \hspace{-25pt}
    \subfloat[]{
  \includegraphics[width=0.42\textwidth]{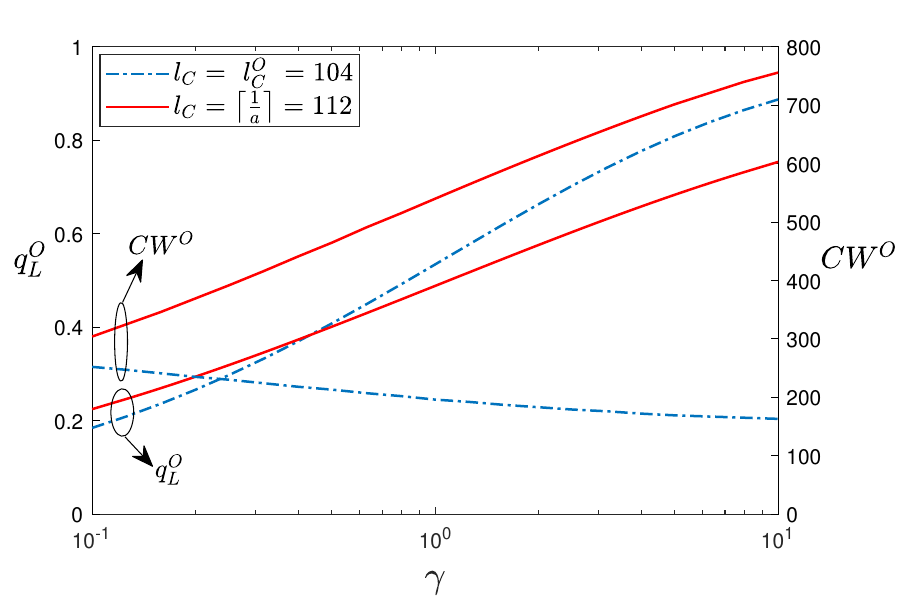}
   \label{fig:opti_CW}}
  \caption{(a)~Total network throughput $\hat \lambda ^{total}_{out }$ with optimal backoff window $CW^O$ and optimal transmission probability $q_L^O$ against desired throughput proportion $\gamma$. $l_C=\left \{ 40,112,224,l_C^O \right \},\ a=0.009,\ n_W=20$. (b)~Optimal backoff window $CW^O$ and optimal transmission probability $q_L^O$ against desired throughput proportion $\gamma$.}
  \vspace{-20pt}
\end{figure*}

\begin{figure}[ht!]
  \centering
  \includegraphics[width=0.4\textwidth]{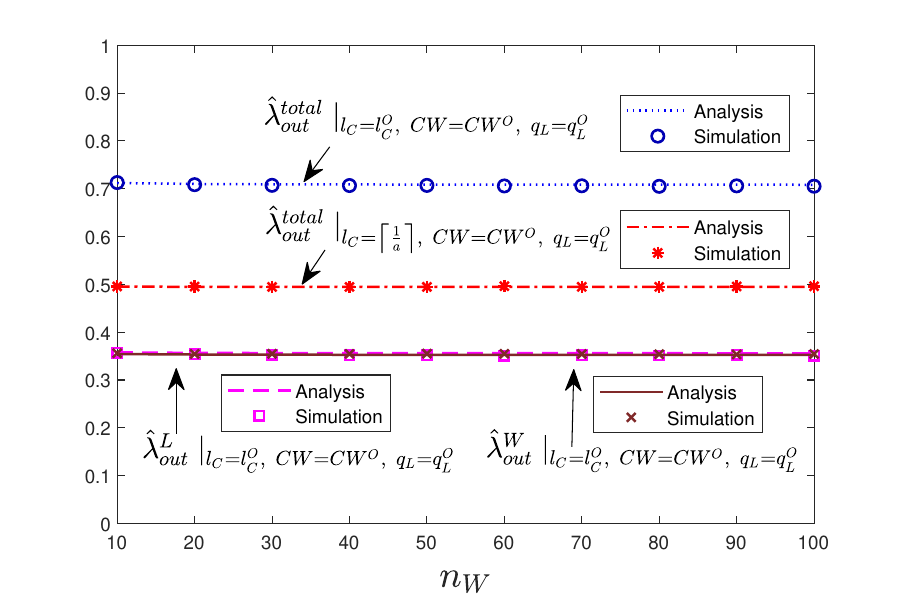}
  \caption{With the optimal configuration and $\gamma=1$, total network throughput $\hat \lambda ^{total}_{out }$, LTE-U throughput $\hat \lambda ^{L}_{out }$ and WiFi throughput $\hat \lambda ^{W}_{out }$ against WiFi network size $n_W$.}
    \label{fig:maxthp_vs_nc}
\end{figure}

\begin{table}[tp]
\centering
\caption{\scalebox{1}{P}\scalebox{0.8}{ARAMETER} \scalebox{1}{S}\scalebox{0.8}{ETTING}\cite{LTEstd,IEEEstd}}
\begin{tabular}{|>{\centering\arraybackslash}p{3.5cm}|>{\centering\arraybackslash}p{2.7cm}|}
\hline
\textbf{Parameter} & \textbf{Value} \\ \hline
The length of ON period                     & 1~ms                            \\ 
Mini-slot~$\sigma$                    & 9~$\mu$s                              \\ 
PHY header                               & 20~$\mu$s                             \\ 
ACK                                      & 112~bits+PHY header                       \\ 
SIFS                                     & 16~$\mu$s                             \\ 
Basic rate                               & 6~Mb/s                            
\\ \hline
\end{tabular}
\label{table:param}
\end{table}

\subsection{Coexistence Model}

The scenario where a duty-cycled-based LTE-U network shares the same unlicensed spectrum with a CSMA-based WiFi network is considered in this section, as illustrated in the left part of Fig.~\ref{fig:Aloha_CSMA_imple}. The LTE-U network schedules only downlink transmissions on the unlicensed spectrum, whereas the WiFi network utilizes the spectrum for both uplink and downlink transmissions. As shown in the upper part of Fig.~\ref{fig:frame_structure}, the scheduling of transmissions in duty-cycled LTE-U network is determined by a sequence of alternating ON and OFF periods, where transmission occurs during ON periods and ceases during OFF periods. Each ON period lasts for 1 ms, which equals the duration of an LTE subframe. Since WiFi nodes are limited to initiate transmission during the OFF period,
it is stipulated by the LTE-U standard \cite{LTEstd} that the minimum OFF period duration must not be shorter than an ON period duration, to protect the throughput performance of WiFi networks. The time axis of the LTE-U network can be segmented into equal intervals of unit duration which equals the length of the ON period, i.e., 1 ms. Randomizing the length of each OFF period can significantly enhance both the fairness and throughput performance of coexisting networks~\cite{9755072}. Fig.~\ref{fig:random_OFF} illustrates transmission scheduling for the LTE-U eNodeB~(eNB) following a randomized OFF period duration. To achieve randomization of an OFF period duration, the LTE-U eNB initiates transmission with a specific probability $q_L$ at the start of each time slot. Consequently, the OFF period duration varies randomly.

The WiFi network adopts the CSMA-based DCF to access the channel. The IEEE 802.11 standard~\cite{IEEEstd} stipulates that each mini-slot~$\sigma$ lasts for $9~\mu\text{s}$. Before each WiFi node transmits, it randomly chooses a backoff counter from $\left\{0,1,\dots,CW_{i-1}\right\}$, where $CW_i=CW\cdot 2^{\text{min}\left ( i,K \right )}$ is the backoff window size, which doubles with each collision encountered by the node until the collision counter $i$ achieves the maximum backoff stage $K$. 
Whenever the channel is idle, the backoff counter is decremented by one. The transmission will be initiated when the backoff counter reaches zero. As illustrated in Fig.~\ref{fig:frame_structure}, the transmission time of a successful packet is set to $l_W$ mini-slots~(or $9\cdot l_W~\mu\text{s}$), which is dictated by the size of the payload and associated overhead.  Note that as the failed packet transmission lacks SIFS and ACK frame, its duration is less than the duration of the successful one. With the system parameters adopted in the IEEE 802.11 standard~\cite{IEEEstd}, which are provided in Table~\ref{table:param}, we can obtain that transmitting ACKs frame and SIFS requires only $\frac{16\mu\mathrm{s} +20\mu\mathrm{s}+112 / 6 \mu\mathrm{s}}{9\mu\mathrm{s}} =6$ mini-slots, which is negligible compared to the packet payload length. Here, the packet transmission time of WiFi nodes $l_W$ significantly impacts the performance of coexisting networks and should be carefully tuned.

\subsection{Performance Optimization of LTE-U and WiFi}

A WiFi network consisting of $n_W$ nodes, operating alongside an LTE-U eNB following a randomized OFF period duration within the same spectrum, can be considered as the cohabitation between a $n_W$-node CSMA network and a single-node Aloha network. Specifically, the LTE-U with randomized OFF periods is a one-node Aloha network, as the LTE-U eNB transmits at the beginning of each time slot with probability $q_L$. Therefore, the analysis in Section \ref{sect:4} can be directly applied to the coexisting LTE-U and WiFi networks. In particular, 
given that each mini-slot and LTE-U ON period last for $9~\mu \text{s}$ and $1~\text{ms}$, respectively, we can obtain that $a=0.009$ and the time slot lasts for $\left \lceil \frac{1}{a} \right \rceil =112$ mini-slots. For the Aloha node, its transmission probability is $q_A=q_L$. Regarding the CSMA nodes, the lengths for both successful and failed packet transmission are $l_C=l_W$ mini-slots. The maximum backoff stage $K$ is set to $0$, and the backoff window size is $CW$. Then the transmission probability of CSMA nodes is $q_C=\frac{2}{CW-1}$ and $\rho_C=(1-\frac{2}{CW-1})^{n_W}$.

To validate the analysis, event-driven simulations are conducted. In simulations, there are $n_W$ WiFi nodes that coexist with an LTE-U eNB. The buffer of each node is always not empty. For the LTE-U eNB, it initiates transmission with a certain probability $q_L$ at the start of each time slot, and the transmission lasts for one time slot~(112 mini-slots). For the WiFi node, it randomly chooses a backoff counter from $\left\{0,1,\dots,CW\right\}$ and senses the channel at the beginning of each mini-slot. Once it senses the channel idle, its backoff counter is decremented by one. It will initiate transmission when the backoff counter reaches zero. Based on the parameters setting in Table~\ref{table:param}, the successful and failed transmission of WiFi nodes lasts for $l_W$ and $l_W-6$ mini-slots, respectively. The simulation results of the LTE-U throughput $\hat \lambda ^{L}_{out}$ and WiFi throughput $\hat \lambda ^{W}_{out}$ are obtained by calculating $\frac{N_L\cdot112}{T}$ and $\frac{N_W\cdot l_w}{T}$, where $N_L$ and $N_W$ represent the number of successful transmissions of the LTE-U eNB and WiFi nodes over a long time period, i.e., each simulation run lasts for $T=10^8$ mini-slots, respectively. The simulations adopt the classic collision model, i.e., a packet transmission is successful if and only if there is no overlap in the packet transmission.

Section \ref{sect:4} demonstrates that to optimize the coexisting network performance, the network parameters, particularly the packet transmission time of CSMA nodes $l_C$, should be carefully tuned. We numerically search for the optimal $l_C^O$. Fig.~\ref{fig:maxthp_vs_gamma_lc} illustrates the total network throughput with optimal backoff window $CW^O$ and optimal transmission probability $q_L^O$ versus the desired throughput proportion $\gamma$. As Fig.~\ref{fig:maxthp_vs_gamma_lc} shows, when $0.1\leq\gamma\leq10$, $l_C^O$ equals 104, which is slightly less than $\left \lceil \frac{1}{a} \right \rceil$. We can observe that the throughput performance is notably better when $l_C=l_C^O=104$, compared to when $l_C$ is equal to $\left \lceil \frac{1}{a} \right \rceil$, i.e., $l_C=112$, which is the $l_C$ setting considered in \cite{9755072}. 

Fig.~\ref{fig:opti_CW} further illustrates the optimal backoff window size $CW^O$ of WiFi nodes and the optimal transmission probability $q_L^O$ in two packet transmission time settings: $l_C=\left \lceil \frac{1}{a} \right \rceil=112$ and $l_C=l_C^O=104$. In both packet transmission time settings, the optimal transmission probability $q_L^O$ steadily increases with the growing desired throughput proportion $\gamma$. Regarding the optimal backoff window, the two settings exhibit different trends. When $l_C=\left \lceil \frac{1}{a} \right \rceil$, the optimal backoff window size $CW^O$ increases as $\gamma$ grows. In contrast, when $l_C=l_C^O$, the optimal backoff window size $CW^O$ decreases as $\gamma$ grows, indicating that more transmission opportunities should be assigned to WiFi nodes.

The reason for different trends is that  WiFi nodes need to tune their backoff window to mitigate interference with the LTE network when $\gamma$ is large, i.e., more priority is given to the LTE-U network. When $l_C=l_C^O$, WiFi nodes can slightly increase the transmission probability to initiate and conclude transmissions promptly within an OFF period. In this way, WiFi nodes can avoid interference with potential transmissions from the LTE-U eNB in the next time slot.  Conversely, for $l_C=\left \lceil \frac{1}{a} \right \rceil$, transmissions of WiFi nodes are unlikely to initiate and conclude within an OFF period. Therefore, WiFi nodes have to decline the transmission probability, at the cost of the WiFi network throughput, to reduce interference with the LTE network. 

Fig.~\ref{fig:maxthp_vs_nc} illustrates the throughput performance of the coexisting network with the optimal setting, as the number of WiFi nodes varies from 10 to 100, in the scenario where $\gamma=1$. In this scenario, it is desired that two networks reach an identical throughput. We can observe that, under the optimal configuration, identical throughput for LTE-U and WiFi is ensured, irrespective of changes in the WiFi network sizes. Furthermore, the total network throughput with the optimal $l_C^O$ is significantly higher than the total network throughput with $l_C=\left \lceil \frac{1}{a} \right \rceil$. 

\begin{figure*}[t]
  \centering
  \subfloat[]{
  \includegraphics[width=0.40\textwidth]{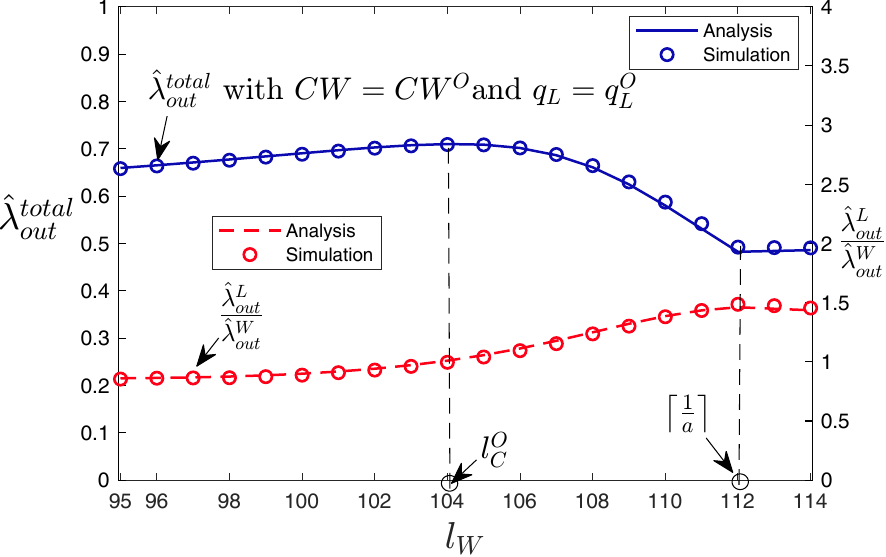}
  \label{fig:wrong_lc}}
    \subfloat[]{
  \includegraphics[width=0.40\textwidth]{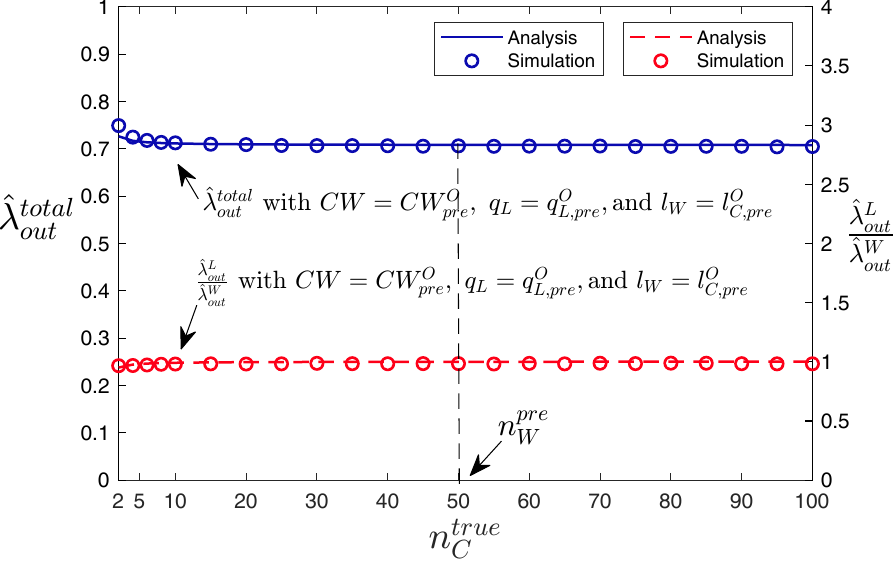}
   \label{fig:wrong_nc}}
  \caption{Illustration of the network performance with (a) deviated packet transmission times of WiFi nodes and (b) the previous number of CSMA nodes. $\gamma=1$. }
   \label{fig:wrong}
   \vspace{-20pt}
\end{figure*}

\subsection{Optimal Configuration Implementation of LTE-U and WiFi}
\label{sect:imple}
This subsection demonstrates how to implement the optimal configuration in the practical coexisting network where an LTE-U eNB coexists with an $n_W$-nodes WiFi network. The implementation process is shown in the right part of Fig.~\ref{fig:Aloha_CSMA_imple}. Searching for the optimal configuration requires the following parameters: the packet transmission time of Aloha nodes $\frac{1}{a}$, the network size, and the desired throughput proportion $\gamma$. The packet transmission time of Aloha nodes can be obtained according to the standard-specified duration of a mini-slot $\sigma=9~\mu \mathrm{s}$ and an LTE-U ON period, i.e., $1~\mathrm{ms}$. For the network size and the desired throughput proportion $\gamma$, the access point~(AP) of the WiFi network and the LTE-U eNB need to first reach a consensus on $\gamma$, and exchange information on the network size. Then the AP and the LTE-U eNB search for the optimal $l_C^O$, $\rho_C^O$, and $\rho_A^O$. Subsequently, the LTE eNB adjusts its transmission probability to $q_L^O$ based on $\rho_A^O$, where $q_L^O=1-\rho_A^O$. The AP calculates the optimal backoff window $CW^O$ based on the WiFi network size $n_W$ and $\rho_C^O$, where $CW^O=\left \lceil \frac{2}{1-(\rho_C^O)^{1/n_W}} +1 \right \rceil$. Then the AP broadcasts the beacon frame which contains the optimal backoff parameters and optimal $l_C^O$ to all WiFi nodes periodically. Each WiFi node updates its configuration upon receiving the beacon frame.

\subsection{Robustness analysis}
This subsection demonstrates the robustness of the proposed implementation. In practical networks, fluctuations in data transmission at the application layer can cause the actual packet transmission time of WiFi nodes $l_W$ to deviate from the obtained optimal packet transmission time $l_C^O$. However, WiFi nodes and LTE-U eNB still contend for the channel using the optimal access parameters $CW^O$ and $q_L^O$. Fig.~\ref{fig:wrong_lc} presents the simulation results of network performance with deviated packet transmission times of WiFi nodes. It is observed that total network throughput and throughput ratio curves remain relatively flat when $l_W$ is less than $l_C^O$, but become steep when $l_C$ is greater than $l_C^O$ yet remains less than $\left \lceil \frac{1}{a} \right \rceil$~(The length of ON period of LTE-U eNB). This indicates that a slight decrease in $l_W$ from $l_C^O$ has minimal impact on total network throughput or fairness. whereas a small increase in $l_W$ beyond $l_C^O$ leads to a sharp decline in both network performance and fairness.

Due to node mobility,  WiFi nodes may enter and exit the network, resulting in changes in WiFi network size. The AP and the LTE-U eNB calculate the optimal transmission parameters $CW^O_{pre}$, $q^O_{L,pre}$, and $l^O_{C,pre}$ based on the previous number of WiFi nodes $n_W^{pre}$. Then the AP  broadcasts the optimal transmission parameter to all WiFi nodes. However, the actual number of WiFi nodes is given by $n_W^{true}$. Fig.~\ref{fig:wrong_nc} illustrates the network performance under this scenario. We can observe that using a previous number of WiFi nodes does not significantly affect network performance.

\section{Conclusion}
\label{sect:6}

In this paper, a novel dual-channel analytical framework is established for ensuring harmonious cohabitation between slotted Aloha and CSMA networks. The network throughputs of Aloha and CSMA are both derived as explicit expressions of the packet transmission
time of CSMA nodes $l_C$, network sizes, and transmission
probabilities. It is found that the throughput performance of the coexisting network closely hinges on the packet transmission time of CSMA nodes $l_C$. To enhance the total network throughput and ensure fair cohabitation, the total network throughput is further optimized subject to a desired throughput proportion by jointly tuning the packet transmission time of CSMA nodes and transmission probabilities of both networks. The optimization results indicate that for a broad range of desired throughput proportions and Aloha network sizes, it is optimal to set the packet transmission time of CSMA nodes slightly less than that of Aloha nodes.

This analysis provides critical guidance on the harmonious cohabitation between practical Aloha-based and CSMA-based networks, such as LTE-U networks coexisting with WiFi networks. By applying the proposed analytical framework, the optimal system configuration of cohabitation between LTE-U and WiFi networks is obtained. The simulation results show that the optimal system configuration guarantees throughput fairness between two networks and substantially enhances the total network throughput in comparison to the default configuration.

\appendices
\section{The Step of Transformation of~(\ref{formula2.3}) }
\label{app:trans}
To transform equation~(\ref{formula2.3}) into an equation set that only includes the limiting state probabilities of idle states~$\tilde{\pi}_{(\mathrm{I}^{(A)},\mathrm{I}^{(C)},D)}$, we need to use $\tilde{\pi}_{(\mathrm{I}^{(A)},\mathrm{I}^{(C)},D)}$ to replace the limiting state probabilities of busy states, i.e., $\tilde{\pi}_{(\mathrm{B}^{(A)},\mathrm{I}^{(C)},D)}$, $\tilde{\pi}_{(\mathrm{I}^{(A)},\mathrm{B}^{(C)},D)}$ and $\tilde{\pi}_{(\mathrm{B}^{(A)},\mathrm{B}^{(C)},D)}$. The following are the detailed transformation steps.

Three kinds of busy states need to be transformed :~$(\mathrm{I}^{(A)},\mathrm{B}^{(C)},D)$, $(\mathrm{B}^{(A)},\mathrm{B}^{(C)},D)$, and $(\mathrm{B}^{(A)},\mathrm{I}^{(C)},D)$.
The first step is to transform the limiting state probabilities of the States $(\mathrm{I}^{(A)},\mathrm{B}^{(C)},D)$ and the States $(\mathrm{B}^{(A)},\mathrm{B}^{(C)},D)$. The relationship between  $\pi_{(\mathrm{I}^{(A)},\mathrm{B}^{(C)},D)}$, $\pi_{(\mathrm{B}^{(A)},\mathrm{B}^{(C)},D)}$, and $\pi_{(\mathrm{I}^{(A)},\mathrm{I}^{(C)},D)}$ is summarized as follows. 
\begin{mys}
\begin{align}
&\hspace{-4.5mm} \begin{aligned}
&\pi_{(\mathrm{I}^{(A)},\mathrm{B}^{(C)},D)}=\\
&\begin{cases} (1-\rho_C) \cdot \pi_{(\mathrm{I}^{(A)},\mathrm{I}^{(C)},D-1)},&D>0,\\ 
 \rho_A\cdot (1-\rho_C)\cdot \pi_{(\mathrm{I}^{(A)},\mathrm{I}^{(C)},\frac{1}{a}+D-1-\frac{\left \lfloor -Da \right \rfloor }{a} ),}{}  &D\le0. \\\end{cases}\\  
 \label{eq:trans_1}
 \end{aligned}\\
 &\hspace{-4.5mm} \begin{aligned}
&\pi_{(\mathrm{B}^{(A)},\mathrm{B}^{(C)},D)}=\\
&(1-\rho_A)\cdot (1-\rho_C)\cdot \pi_{(\mathrm{I}^{(A)},\mathrm{I}^{(C)},\frac{1}{a}+D-1-\frac{\left \lfloor -Da \right \rfloor }{a} ),}{}  &D\le0.
\label{eq:trans_2}
  \end{aligned}
\end{align}    
\end{mys} 
\hspace{-1.7mm}The above relationship can be obtained from the CSMA sensing mechanism. 

The second step is to transform the remaining limiting state probabilities of the busy states $(\mathrm{B}^{(A)},\mathrm{I}^{(C)},D)$. $\tilde{\pi}_{(\mathrm{B}^{(A)},\mathrm{I}^{(C)},k)}$ can be expressed as \footnote{For brevity, we adopt the following short notations: 
$\mathrm{Pr}\left\{\boldsymbol{\mu } \right\}=\mathrm{Pr} \{ \mathrm{the\ coexisting\ network\ state\ is\ \boldsymbol{\mu} } \}$, where $\boldsymbol{\mu}\in \mathbb{S}$, and $\mathrm{Pr}\left\{ (t)_i\right\}=\mathrm{Pr} \{ \mathrm{the\ coexisting\ network\ is\ at\ }(t)_i \}$, where $i=1,2,\dots,\frac{1}{a}$. } 
\begin{equation}
\label{qiuhe_1}
\begin{aligned}
\tilde{\pi}_{(\mathrm{B}^{(A)},\mathrm{I}^{(C)},k)}&=a- \sum_{\substack{\boldsymbol{\mu} \in \mathbb{S}, \\ \boldsymbol{\mu} \neq (\mathrm{B}^{(A)},\mathrm{I}^{(C)},k)}}^{} \mathrm{Pr} \{ \boldsymbol{\mu}  ,\ (t)_{k+1}   \}.
\end{aligned}
\end{equation}
where each non-zero term in the summation on the right-hand side of the~(\ref{qiuhe_1}), can be represented by $\tilde{\pi}_{(\mathrm{I}^{(A)},\mathrm{I}^{(C)},D)}$.~(\ref{qiuhe_1}) can be derived using conditional probability.
With~(\ref{qiuhe_1}), $\tilde{\pi}_{(\mathrm{B}^{(A)},\mathrm{I}^{(C)},D)}$ can also be represented by $\tilde{\pi}_{(\mathrm{I}^{(A)},\mathrm{I}^{(C)},D)}$.

Using the above two steps, equation~(\ref{formula2.3}) can be transformed into an equation set that only includes the limiting state probabilities of idle states~$\tilde{\pi}_{(\mathrm{I}^{(A)},\mathrm{I}^{(C)},D)}$.



\bibliographystyle{IEEEtran}
\bibliography{IEEEabrv,template}
%


\vfill


\end{document}